\documentclass{emulateapj}
\usepackage{epsfig} 
\usepackage{amsmath}
\usepackage{amssymb}
\usepackage{natbib}
\usepackage{graphicx}

\shorttitle{Gravitationally unstable gas-rich galactic disks}
\shortauthors{Aumer, Burkert, Johansson, Genzel}
\begin{document}

\title{The structure of gravitationally unstable gas-rich disk galaxies }

\author{Michael Aumer$^{1,2}$, Andreas Burkert$^1$, Peter H. Johansson$^1$, Reinhard Genzel$^{3,4}$}
\affil{$^1$ Universit\"ats-Sternwarte M\"unchen, Scheinerstr.\ 1, D-81679 M\"unchen, Germany;\\
$^2$ Max-Planck-Institut f\"ur Astrophysik (MPA), Karl-Schwarzschild-Str. 1, 85748 Garching, Germany;\\
$^3$ UC Berkeley Department of Physics, Berkeley, CA 94720, USA;\\
$^4$ Max-Planck-Institut f\"ur extraterrestrische Physik (MPE), Giessenbachstr. 1, 85748 Garching, Germany;\\
 \texttt{aumer@usm.lmu.de, burkert@usm.lmu.de, pjohan@usm.lmu.de, genzel@mpe.mpg.de} \\}

\begin{abstract}

We use a series of idealized, numerical SPH simulations to study the formation and evolution of galactic, 
gas-rich disks forming from gas infall within dark matter halos. The temperature and
density structure of the gas is varied in order to differentiate between (i) simultaneous 
gas infall at a large range of radii and (ii) the inside-out build-up of a disk. 
In all cases, the disks go through phases of ring formation, gravitational instability and 
break-up into massive clumps. Ring formation can be enhanced by a focal point effect.
The position of the ring is determined by the angular momentum
distribution of the material it forms from. We study the ring and clump morphologies, 
the characteristic properties of the resulting velocity dispersion field and the effect of star formation.
In the early phases, gas accretion leads to a high vertical velocity dispersion.
 We find that the disk fragmentation by gravitational instability and 
the subsequent clump-clump interactions drive high velocity dispersions mainly in the plane of the disk while
at the same time the vertical velocity dispersion dissipates. The result is a strong
variation of the line-of-sight velocity dispersion with inclination angle.
For a face-on view, clumps appear as minima in the (vertical) dispersion, whereas for
a more edge-on view, they tend to correspond to maxima.
There exists observational evidence of a systematic variation of the
velocity dispersion with inclination angle in high-redshift disks, which could be partly explained
by our simulation results. Additional energetic sources to drive velocity dispersion that
are not included in our models are also expected to contribute to the observational
results.

\end{abstract}


\section{Introduction}

Recent observations at redshift $z\sim 2$, close to the peak of the cosmic star
formation history, have revealed a significant population of massive (baryonic 
mass $M_{\rm bar}\sim 10^{11} M_{\odot}$), extended ($R \sim 6\; \rm{kpc}$) disk galaxies with
star formation rates of the order of $\rm{SFR} \sim 100\; M_{\odot}\;\rm{yr^{-1}}$
(e.g. \citealp{2006ApJ...645.1062F,2009ApJ...706.1364F}).
The corresponding observed gas velocity dispersions are very high, $\sigma \sim 40 - 90\, \rm{km \; s^{-1}}$
and the rotation-to-dispersion ratios $V/\sigma \sim 2-6$ thus relatively low (e.g. \citealp{2009ApJ...697..115C, 
2006Natur.442..786G,2008ApJ...687...59G}). In addition, the galaxies are gas-rich with
typical gas fractions of $\sim50\%$ (\citealp{2009arXiv0911.2776D, 2010arXiv1002.2149T}). Star formation
within them takes place in a few giant clumps with typical masses of
$M_{\rm clump}\sim 10^9 M_{\odot}$
and radii of $R_{\rm clump}\sim 1 \ \rm{kpc}$ \citep{2009ApJ...692...12E}.  In some of the sources 
rings have been detected, suggesting an evolutionary sequence from rings to clumps 
to bulges on a timescale of $t\lesssim 1 \ \rm{Gyr}$ (\citealp{2008ApJ...687...59G}).

The properties of this observed $ z \sim 2$ galaxy population are very different from present-day
disk galaxies. Typical disk galaxies, such as the Milky Way, display smooth
surface density profiles with no evidence for supermassive clumps.
The difference to the $ z \sim 2$ population can be illustrated 
by the typical parameters for the local disk population of
$\rm{SFR} \sim 1 \ M_{\odot}\;\rm{yr^{-1}}$ (\citealp{2010ApJ...710L..11R} and references therein) and
$V/\sigma \sim 20$ (\citealp{2006ApJ...638..797D}). Star formation takes place in
giant molecular clouds with typical masses of $M_{\rm GMC}\sim 10^6 M_{\odot}$ and has been
constant within a factor of a few for the last $t\sim 10 \ \rm Gyr$ (\citealp{2009MNRAS.397.1286A}).
Associations of molecular clouds can have masses as high as a few $10^7 M_{\odot}$ (e.g. \citealp{1990ApJ...349L..43R}), which is the
largest scale of fragmentation in $z\sim 0$ galaxies.
However, the thick disk components of low-redshift disks  
exhibit stellar ages consistent with a formation at $z \sim 2$ and high velocity
dispersions (e.g. \citealp{2009MNRAS.399.1145S}). It has been proposed that
the thick disks have formed in massive, clumpy high-redshift disks with 
high SFRs \citep{2006ApJ...645.1062F, 2006Natur.442..786G,2007ApJ...670..237B,2009ApJ...707L...1B} . 

Observations indicate that only a smaller fraction $(\sim 1/3)$ of the $z\sim
2$ galaxies are obvious major mergers 
\citep{2006ApJ...645.1062F,2009ApJ...706.1364F,2008ApJ...682..231S}. Thus,
the high star formation rates and high velocity dispersions seem not to be
merger-induced in the majority of the sources. Instead, gravitational
instability of differentially rotating disks with high surface densities
(\citealp{1964ApJ...139.1217T}) has been identified as a key process for the
fragmentation of $z\sim2$ disk galaxies, which are then likely to evolve into 
bulge-dominated systems (\citealp{1999ApJ...514...77N,2004A&A...413..547I}).
Lately, \citet{2010arXiv1001.4732R} extended the classical Toomre instability criterion 
to turbulent disks which show a rich variety of global instabilities, providing
a theoretical framework for gas-rich high-z disk evolution.

Clumpy and turbulent disks resembling the $z \sim 2$ galaxies have recently also been produced
in cosmological simulations of galaxy formation (\citealp{2009MNRAS.397L..64A,2009arXiv0907.3271C}).
As argued by the authors, key aspects in the success of these simulations are a high spatial resolution
and a cooling description that follows the thermal evolution of the gas below
$T<10^{4} \ \rm K$ allowing a multi-phase ISM to develop.
In these simulations, the high gas fractions and star-formation rates are powered by cold streams
of gas moving along the cosmic web (\citealp{2005MNRAS.363....2K,2009ApJ...703..785D, 2009ApJ...697L..38J}).
However, the physical source of the large observed irregular motions is
still unclear. Several scenarios have been proposed in the literature
including accretion (e.g. \citealp{2006ApJ...645.1062F, 2006Natur.442..786G,2009arXiv0912.0996E,2009arXiv0912.0288K}), stellar
radiative feedback (\citealp{2010ApJ...709..191M,2010arXiv1001.0765K}) and disk
self-gravity (e.g. \citealp{2004A&A...413..547I,2007ApJ...670..237B,2009arXiv0907.4777B}).

In this paper we present SPH simulations of idealized disk formation by gas infall via a cooling flow
within a galactic halo. We analyze the evolution of the morphology and the
turbulent structures in the forming disks. Both models with and without star
formation and SN feedback and with different initial temperature and density profiles are presented.
We demonstrate that ring formation and subsequent gravitational fragmentation
occurs in all of the scenarios. In addition, we analyze the structure of
velocity dispersion created by accretion and disk self-gravity and finally discuss the
efficiency of stellar disk heating in our simulations.

We describe the setup of our simulations in Section 2, present our results in Section 3
and discuss them in Section 4.

\section{Simulations}

The simulations were performed using the TreeSPH-code GADGET-2
\citep{2005MNRAS.364.1105S} on the local Altix 3700 Bx2 machine.
The code includes the standard radiative cooling rates for an optically thin, primordial composition of
hydrogen and helium in ionization equilibrium following \citet{1996ApJS..105...19K}.
Their figure 1 displays the cooling rate as a function of temperature. The cooling curve features a cutoff 
below $T\sim10^4 \rm K$ and two peaks at $T\sim10^{4.3} \rm K$ and $T\sim10^{5.0} \rm K$
resulting from collisional excitation of H and He, respectively.
We included a spatially uniform time-independent UV background, appropriate for $z=2$
\citep{1996ApJ...461...20H}. The abundances of the different ionic H and He species were computed 
by solving the network of equilibrium equations self-consistently for a 
specified value of the UV background radiation field. 

The chemical composition of the infalling gas could possibly be pre-enriched.
However, the actual metal content is difficult to estimate, as there are no direct observations
of infalling gas at $z\sim2$ (e.g. \citealp{2010arXiv1003.0679S}) and observed galaxies
and outflows are likely to be significantly more metal-rich than the infalling gas.
Considering the metallicity of tyoical damped Lyman $\alpha$ systems at $z\sim2$, $Z\sim0.1Z_{\odot}$
\citep{1994ApJ...426...79P}, the cooling rates would be enhanced for temperatures
$T\gtrsim10^{4.5}K$ by a factor of a few \citep{1993ApJS...88..253S}. In our simulations, this would
lead to an increase in gas mass flow to the center of the halo. However, due to the large uncertainties
in the actual gas metallicity at $z\sim2$, we rely on the assumption of a primordial chemical composition
in the present study.

Star formation and the associated supernova feedback, when included, are modeled using the prescriptions 
of \citet{2003MNRAS.339..289S}. This sub-resolution model uses a statistical formulation to deal with physics
acting on scales, which are not resolved by simulations of galaxy formation. Above a threshold
density $\rho_{\rm th}$ , each SPH particle is assumed to represent a fluid comprised of cold, condensed clouds
in pressure equilibrium with an ambient hot gas. Such a multiphase treatment of the interstellar
medium is motivated by \citet{1977ApJ...218..148M} (see also \citealp{2006MNRAS.371.1519J}). 
A stellar population is modeled to form from the cold medium on a timescale $t_s$ , which is proportional to
the dynamical timescale of the gas, so that $t_s = t_{s,0} (\rho/\rho_{\rm th} )^{-0.5}$ . 
The energy released by one SNII is typically $10^{51}$ ergs per supernova, which may be expressed as a supernova
temperature $T_{\rm SN}=2 \mu u_{\rm SN}/(3 k_B) \sim 10^8  \rm K$.
The hot phase is required to have a temperature in excess of $10^5 \rm K$,
whereas the cold phase is assumed to be at $T_c \sim 10^3 \rm K$.
The threshold density, $\rho_{\rm th}$ , is determined self-consistently in the model
by requiring that the equation-of-state (EOS) is continuous at the onset of star formation.
Stars form from the cold clouds in regions were the number density $n > n_{\rm th} = 0.128 \rm cm^{-3}$.
The parameters governing the model are set to reproduce the observed Kennicutt relation \citep{1998ARA&A..36..189K},
yielding a star formation timescale $t_{s,0} = 2.1 \rm Gyr$.
The unphysically low star formation density threshold density $n_{\rm th}$ is a consequence
of the relatively low resolution of typical galaxy formation simulations, 
such as ours ($\sim 10^5 M_{\odot}$ per particle).
As has recently been shown \citep{2010Natur.463..203G}, higher density
thresholds are appropriate for higher-resolution simulations of dwarf galaxies ($\sim 10^3 M_{\odot}$ per particle).

Radiation pressure, which might play an important role in high-$z$ star forming galaxies
(see \citealp{2010ApJ...709..191M}), is not included in our simulations.

The setup for our simulations was motivated by \citet{2006MNRAS.370.1612K,2007MNRAS.375...53K}.
We study the evolution of an initially hot baryonic component within a dark matter 
halo with an NFW density profile 
\begin{equation}
\label{NFW}
\rho(R)=\frac{\rho_s}{R/R_s(1+R/Rs)^2}
\end{equation}
\citep{1996ApJ...462..563N}.
The structural parameters of the halo were chosen according to \citet{2009ApJ...707..354Z}.
The virial radius $R_{\rm vir}$ and virial mass $M_{\rm vir}$ are defined
according to the spherical virialization criterion.
The corresponding virial over-density $\Delta_{\rm vir}(z\sim2)\approx 187$ depends on redshift $z$ and
on the assumed cosmology. 
We adopt $M_{\rm vir}=10^{12} M_{\odot}$ as expected for $z\sim2$ star-forming galaxies \citep{2009ApJ...703..785D}.
For this halo mass, $z=2$ and a $\Lambda$CDM cosmology with parameters according to \citet{2009ApJS..180..225H}, 
\citet{2009ApJ...707..354Z} find a concentration parameter of $c=R_{\rm vir}/R_s=4.2$ and a virial 
radius of $R_{\rm{vir}}=108\; \rm{kpc}$, which we adopt. 
We apply a fiducial virial spin parameter of $\lambda_{\rm{DM}}=\frac{j_{\rm{DM}}}{\sqrt{2}VR_{\rm vir}}=0.08$ 
(\citealp{2001ApJ...555..240B}), where $j_{\rm{DM}}$ is the average specific angular momentum of the 
dark matter within $R_{\rm vir}$ and $V=\sqrt{G M_{\rm vir} / R_{\rm vir} } $ is the halo circular velocity 
at $R_{\rm vir}$.  This value for $\lambda_{\rm{DM}}$ is on the high end of the distribution
of spin parameters of CDM halos found by \citet{2001ApJ...555..240B}. However, similarly high spin parameters
have been inferred for extended disk galaxies at $z\sim2$ \citep{2009arXiv0907.4777B}, thus motivating
our choice. 

The halo is set up using a population of $N_{\rm{DM}}=10^6$ collisionless
particles following the description of \citet{2005MNRAS.361..776S} (see also \citealp{2009ApJ...690..802J}).
We truncate the halo at $1.5\; R_{\rm vir}$. The particle mass is $m_{\rm{DM}}=1.34\times10^6 M_{\odot}$ and 
the gravitational softening length is set to $\epsilon_{\rm{DM}}=500\;\rm{pc}$.

A gas component was included with a baryonic-to-dark mass ratio of $15\%$.
The gas component is also truncated at $1.5\; R_{\rm vir}$ .
We use $N_{\rm{bar}}=10^6$ particles, a particle mass $m_{\rm{bar}}=2.01\times10^5 M_{\odot}$ and 
a gravitational softening length of $\epsilon_{\rm{bar}}=250 \ \rm{pc}$ .
The dominant mode of gas accretion onto forming galaxies is a topic still under debate
\citep{2009Natur.457..451D}. Smooth inflows
of cooling gas, that has previously been shock-heated, cold streams and mergers are among the
discussed modes and are all likely to play a role. We intend to keep our models simple in this
context and therefore assume spherically symmetric profiles of density $\rho_{\rm bar}$ and 
temperature $T$. The variation of these profiles allows us to study different scenarios of infall.
In our models, the gas component is initially hot ($T\gtrsim10^6\rm{K}$), with an internal energy structure 
determined by the assumption of hydrostatic equilibrium (cf. \citealp{2006MNRAS.370.1612K}).
The pressure profile is thus
\begin{equation}
\label{HSE}
p(R)=\int_R^\infty \rho_{gas}\;\frac{G\;M_{\rm{tot}}(r)}{r^2}\;dr
\end{equation}
Unlike Kaufmann et al., we do not impose a lower temperature floor, 
as we do not want to prevent the forming disk from fragmenting.
The build-up of the disk is determined by the initial profiles of gas density, internal energy
and angular momentum. We vary the initial conditions as follows:

\begin{itemize}

\item Model A assumes that the gas component is initially well mixed with the dark matter component.
      Thus, the gas density also follows an NFW density profile similarly to the DM component, but with a 
      correspondingly lower density normalization $\rho_s$ (cf. \citealp{2006MNRAS.370.1612K}). 
      The gas is in hydrostatic equilibrium with an initial temperature of the order of $T=10^6 \ \rm K$ 
      within $R<20 \ \rm kpc$. The initial specific angular momentum profile is a power law with $j \propto R$
      (cf. \citealp{2006MNRAS.370.1612K,2001ApJ...555..240B}) , i.e. the rotational velocity 
      is independent of the radius.

\item With Model B we intend to explore significantly different central initial conditions
      and to consequently alter the initial formation stages. We assume a constant gas density 
      within $R<20 \ \rm kpc$, adjusted so that the total mass within this radius is equal to that of Model A,
      $M_{R<20}=2.5\times10^{10}M_{\odot}$. The temperature in this model at $R=20 \
      \rm kpc$ is also of the order of $T=10^6 \ \rm K$. However, due to the assumption of hydrostatic equilibrium the
      temperature increases with decreasing radius resulting in a factor of $\sim5$
      higher temperature in the center, thus creating a 'hot bubble'.
      In the constant density region the angular velocity of the gas is set to a constant value. 
      Outside $R=30\ \rm kpc$ the density and angular momentum profiles are as in Model A with a smooth transition 
      region between 20 and 30 kpc.

\item Model C has identical gas density and angular momentum profiles to Model B, but the condition
      of hydrostatic equilibrium is removed in order to model an accretion history that is intermediate to Models
      A and B. Within $R<15\ \rm kpc$ the temperature is set to a constant value of
      $T\sim3\times10^6\ \rm{K}$, being lower than the central temperature of Model B, but higher than the
      central temperature of Model A. Between 15 and 30 kpc the temperature decreases linearly with radius and
      outside $R>30\ \rm kpc$ the temperature profile is identical to Model B.

\item With Model SF we intend to study the effects of star formation and supernova feedback on the evolution of
      our systems. As has been shown by \citet{2004A&A...413..547I}, the specific implementation of these
      processes has a large influence on the formation and evolution of unstable disks. We therefore rely
      on the generic model of \citet{2003MNRAS.339..289S} to be able to draw general conclusions.
      We choose identical initial conditions to Model B in order to make a direct comparison possible.

\end{itemize}

The spin parameter $\lambda_{\rm{gas}}=\frac{j_{\rm{gas}}}{\sqrt{2}VR_{\rm{vir}}}$,
where $j_{\rm{gas}}$ is the average specific angular momentum of the gas within $R_{\rm{vir}}$
is for all models set to a fiducial value of 0.08, so that $\lambda_{\rm{gas}}=\lambda_{\rm{DM}}$.

\begin{figure}
\centering 
\includegraphics[width=8.5cm]{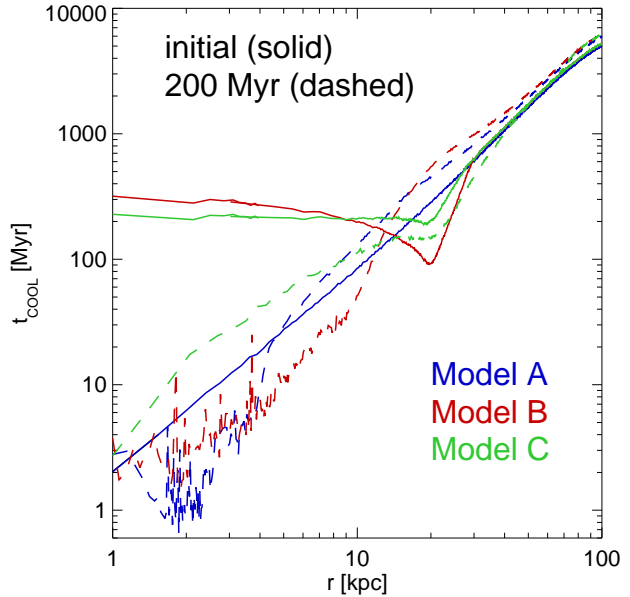}
\caption{Cooling timescales $t_{\rm cool}$ as a function of radius $r$ for Models A (blue),
         B (red) and C (green). The solid lines are for the initial conditions, whereas
         the dotted lines are for simulation time $t=200 \ \rm Myr$.
}
\vspace{0.5cm}
 \label{tcool}
\end{figure}

\section{Results}

Due to cooling the baryonic component evolves out of hydrostatic equilibrium rapidly and begins
to form a disk in the center of the halo. The cooling timescale, $t_{\rm cool} \propto \rho^{-1}$ 
and the temperature, density and angular momentum profiles determine the
evolution and structure of the forming baryonic component.
In Fig. \ref{tcool} we plot the cooling timescale as a function of radius $R$
for the three different initial conditions of Models A/B/C and for the evolved systems
at simulation time $t=200 \ \rm Myr$ in order to illustrate the significant differences between
the three models.

As already noted in \citet{2006MNRAS.370.1612K} thermal instabilities 
(\citealp{2000ApJ...537..270B}) caused by numerical noise in the 
initial conditions lead to the formation of cold, dense clouds 
in the halo which fall onto the disk through a hotter medium.
This however does not affect the fragmentation of our disks
into clumps. The clouds typically consist of a number
of SPH particles $N_{\rm{cloud}}$, which is smaller than 
the number of particles in the smoothing kernel
$N_{\rm{SPH}}=40$ and are thus not resolved. 

\subsection{How the Models differ}

Our different initial conditions have been constructed in order to study 
different early phases of disk growth.

\begin{figure}
\centering 
\includegraphics[width=8.5cm]{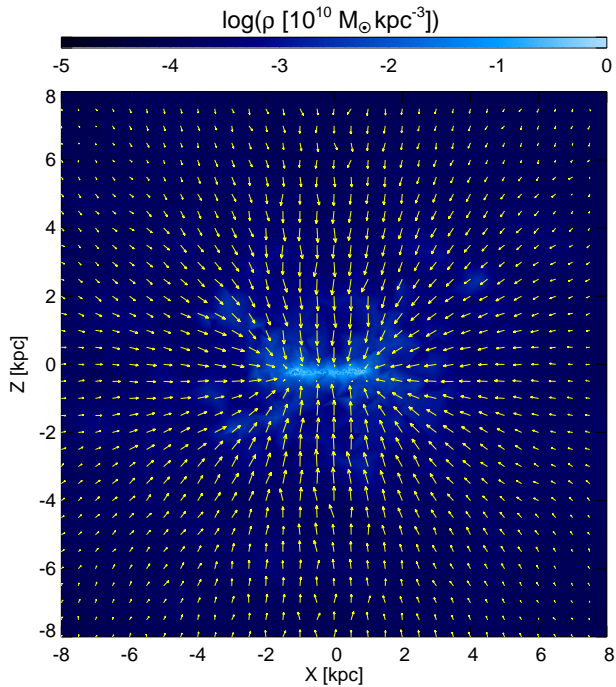}
\caption{A cut perpendicular to the disk region (y=0) of Model A at $t=35 \ \rm Myr$. 
         The color represents density $\rho$ and over-plotted in yellow is the velocity field
         illustrating the focusing effect leading to accretion into the ring
         region. The velocity increases from a few $10\ \rm{km s^{-1}}$ to $\sim160\ \rm{km s^{-1}}$
         near the disk.
}
\vspace{0.5cm}
\label{focus}
\end{figure}

\begin{itemize}

\item In Model A the cooling time increases monotonically as a function of increasing radius. The already 
      relatively dense central gas ($R\lesssim 4 \ \rm kpc$) cools very rapidly
      on a timescale of $t_{\rm{cool}}\lesssim 20\ \rm{Myr}$ (cf. Fig. \ref{tcool}) and forms a 
      rotationally stabilized inner disk. Material within $R\lesssim1\ \rm{kpc}$ of the rotation axis 
      is in centrifugal
      equilibrium and falls vertically onto the disk, whereas the remaining gas
      falls in radially (cf Fig. \ref{focus}). Because of the peaked density profile 
      and correspondingly short collapse timescales, the accretion of 
      material that was originally close to the rotation axis and therefore in centrifugal equilibrium 
      soon becomes unimportant. Now gas with higher angular momentum from the outer regions  begins to
      settle into the outer regions of the disk, leading to an inside-out growth of the disk.
      In general, in this model, accretion predominantly takes place in the outer disk region.

\item According to equation \ref{HSE} the temperature decreases
      with increasing radius in the constant density region of Model B,
      thus the outer regions of the constant density sphere have the
      shortest cooling times $t_{\rm cool}\sim 100 \rm Myr$ (cf. Fig. \ref{tcool}) 
      and the first dense and cold structures are formed here.
      These clouds are gravitationally accelerated inwards,  moving towards the center
      through the hotter inner gas layers, which subsequently also cool and undergo the same process.
      The result is that almost all of the material
      from the initial constant density region settles in a disk at a
      relatively large range of radii ($R \lesssim 5\ \rm{kpc}$)
      within a relatively small time interval ($t\sim 200 - 260\ \rm{Myr}$). The formation 
      phase of the central disk is thus distinctly different from Model A.

\begin{figure*}
\centering 
\vspace{-0.1cm}
\includegraphics[width=16.5cm]{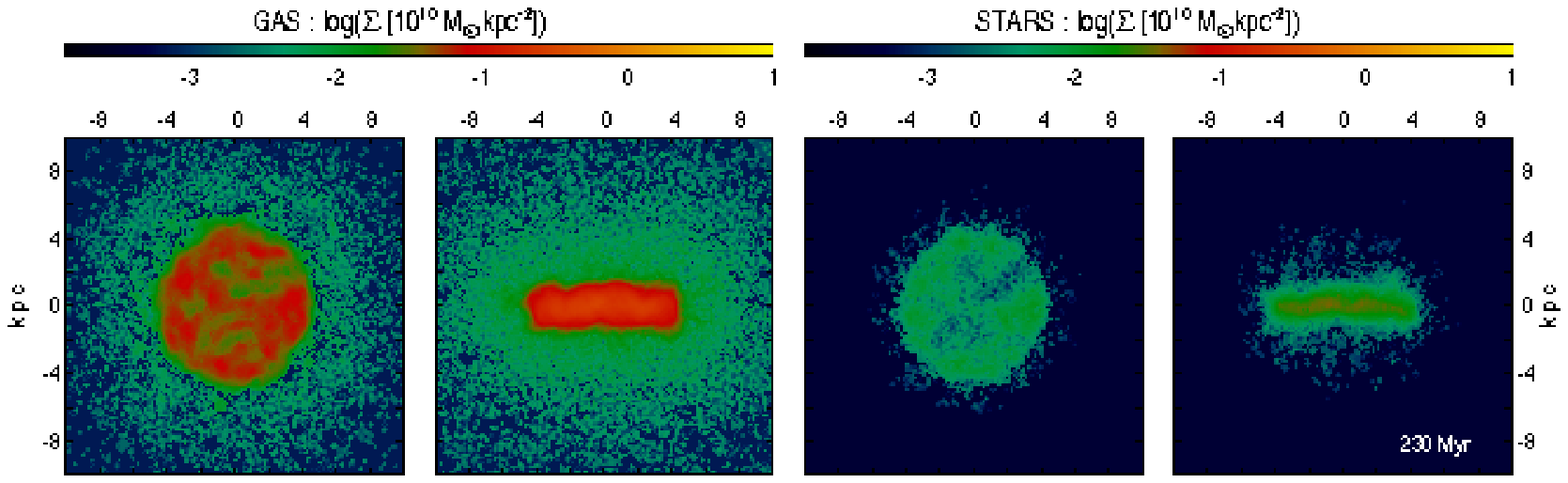}
\vspace{-0.3cm}
\includegraphics[width=16.5cm]{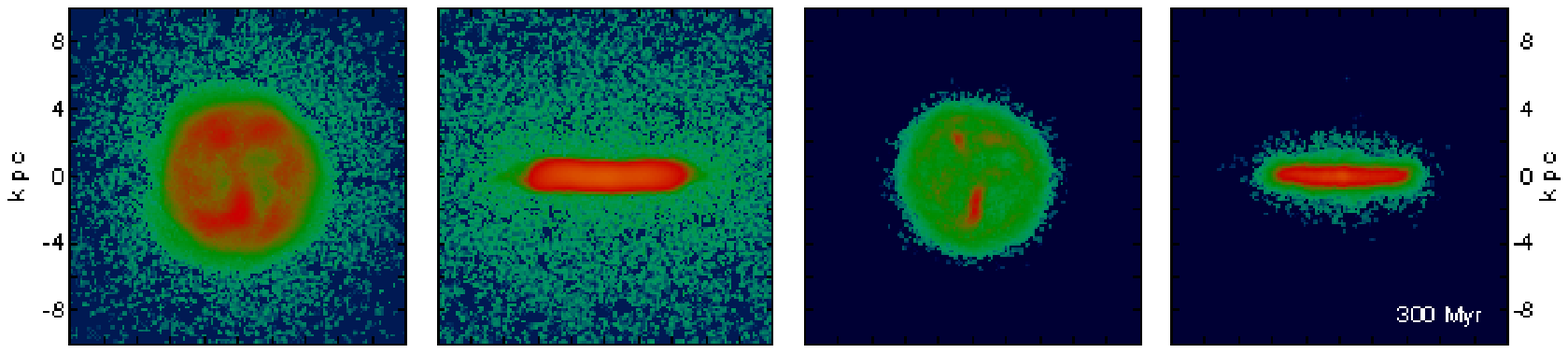}
\vspace{-0.3cm}
\includegraphics[width=16.5cm]{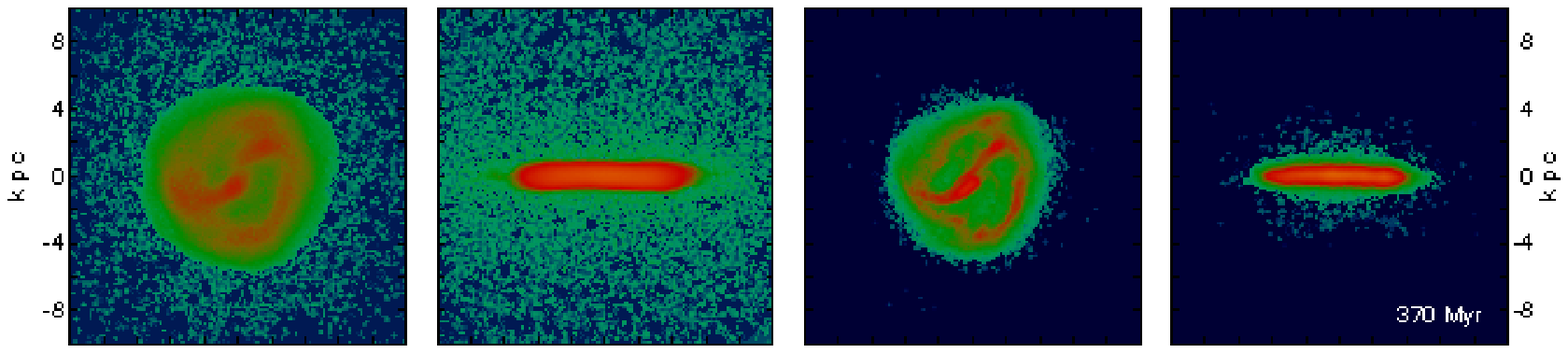}
\vspace{-0.3cm}
\includegraphics[width=16.5cm]{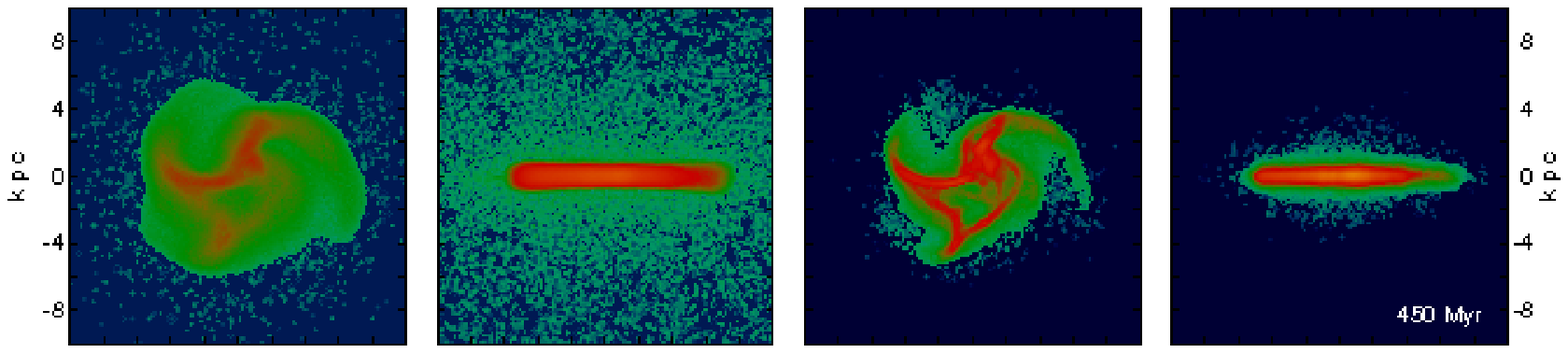}
\vspace{-0.3cm}
\includegraphics[width=16.5cm]{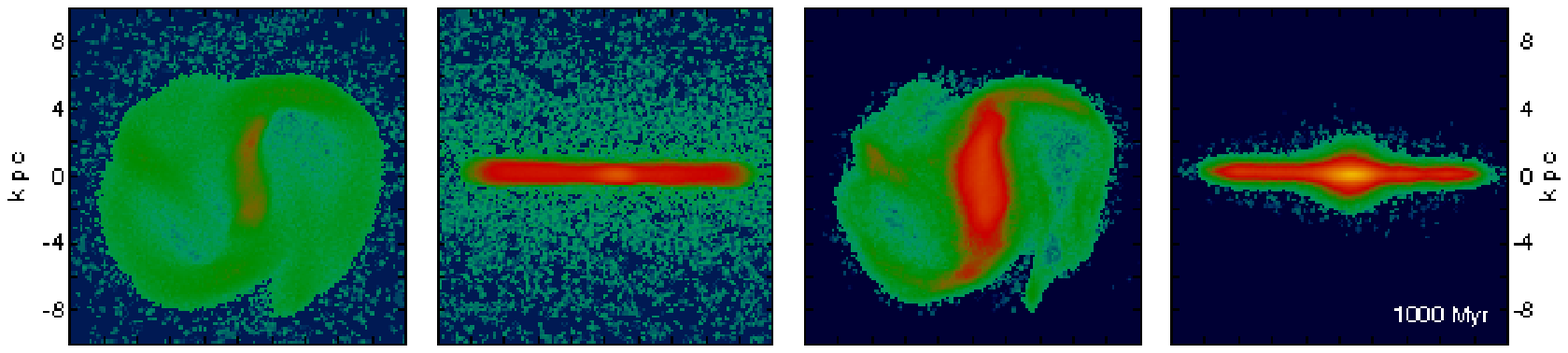}
\vspace{-0.cm}
\includegraphics[width=16.5cm]{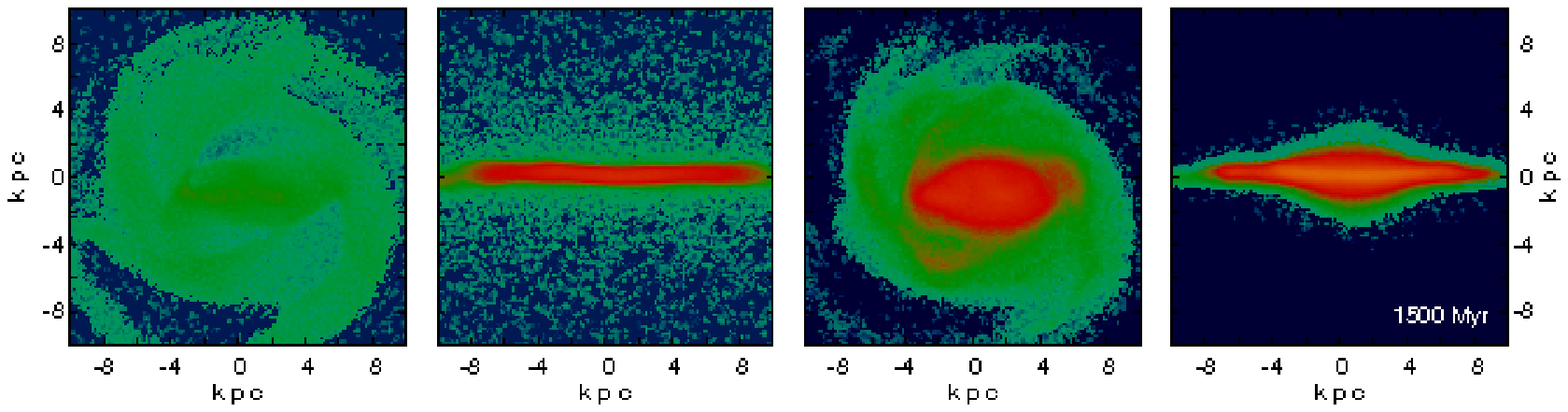}
\caption{The variety of morphologies displayed by Model SF.
         From left to right: Surface density $\Sigma$ maps of: 
         gas face-on, gas edge-on, stars face-on and  stars edge-on.
         From top to bottom: Model SF at 230, 300, 370, 450, 1000, 1500 Myr. }
\label{morph}
\end{figure*}

\item In Model SF star formation sets in after the first denser structures have
      formed due to cooling and before they enter the disk region. However star formation rates in this phase
      are low compared to the gas accretion, leading to an increase in the disk's gas mass. The inclusion of
      feedback increases the effective cooling timescales as the type II supernovae add
      energy to the star-forming gas. The central build-up of the disk
      is thus slightly delayed ($t\gtrsim220 \ \rm{Myr}$) but otherwise similar
      to Model B. The first row of Fig. \ref{morph} displays the face- and edge-on surface
      density maps of Model SF at $t \sim 230 \ \rm{Myr}$.

\begin{figure*}
\centering 
\includegraphics[width=17cm]{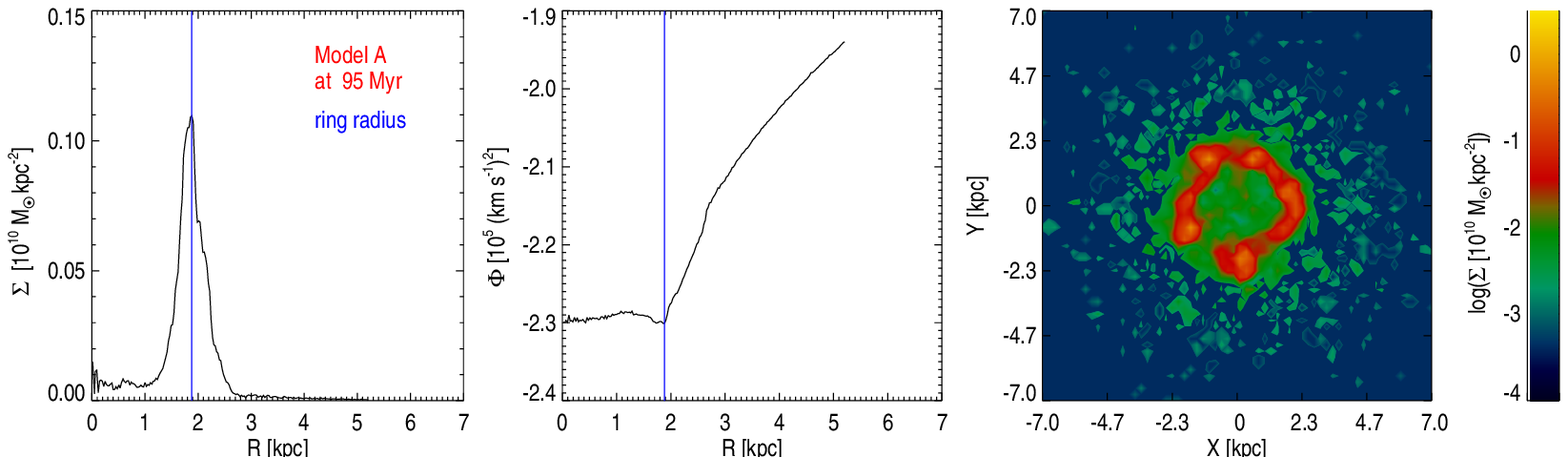}
\includegraphics[width=17cm]{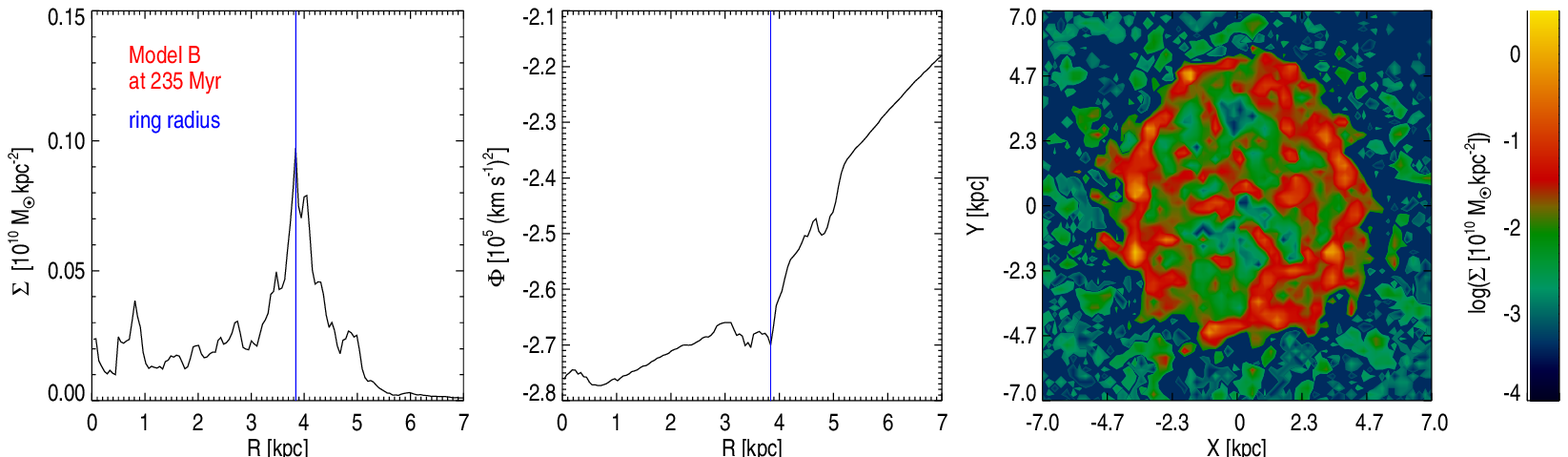}
\includegraphics[width=17cm]{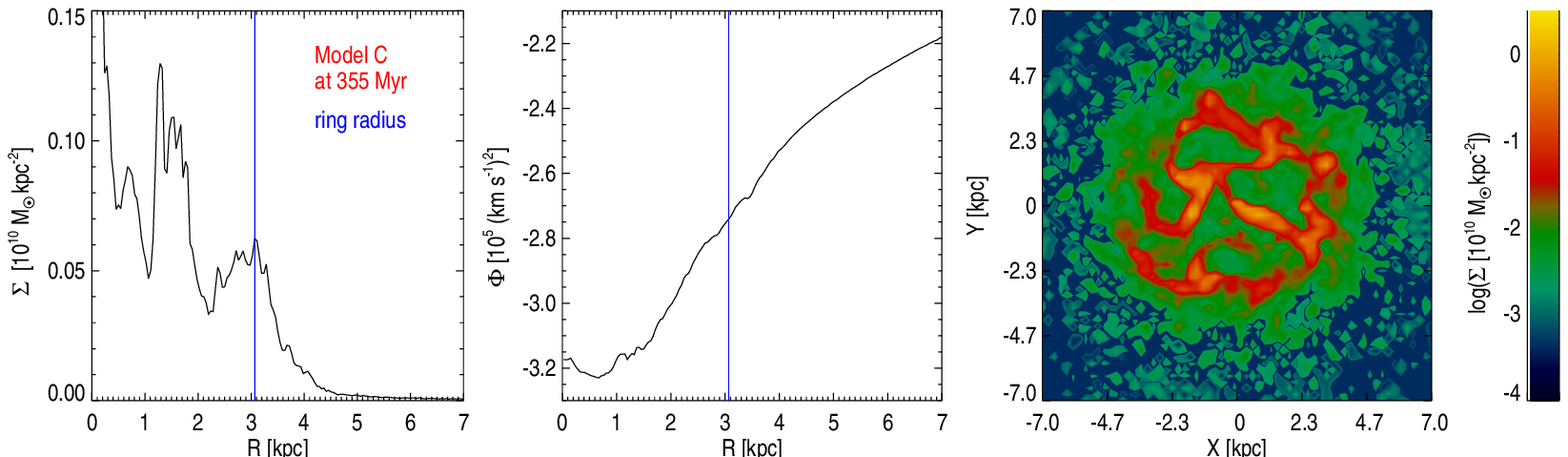}

\caption{From left to right: surface density $\Sigma$ vs disk radius $R$, the azimuthally averaged 
                             gravitational potential $\Phi$ vs disk radius $R$ and
                             a surface density map of the disk regions. The vertical blue
                             lines in the left and central panels mark the position of the ring.
                             From top to bottom: Model A at 95 Myr, Model B at 235 Myr and Model C at 355 Myr.
}
\label{rings}
\end{figure*}

\item Because of the constant density and temperature of the gas within the
      central $R<15\ \rm kpc$ in Model C the cooling time is constant everywhere, $t_{\rm cool}\sim 200 Myr$
      (cf. Fig. \ref{tcool}).
      After a first quasi-static cooling phase, infall starts basically everywhere at the same time.
      The free-fall timescale increases with radius, resulting in an inside-out disk formation starting
      at $t\sim 140 \ \rm Myr$. Compared to Model A the different density
      and angular momentum profiles result in the accretion into the central regions 
      being significant over a much longer timescale (until $t\sim320\ \rm{Myr}$).

\end{itemize}

The dashed lines in Fig. \ref{tcool} show the cooling time profiles for simulation time $t=200 \ \rm Myr$
 for the Models A/B/C. For Models A and B, extended central structures have formed, indicated
 by the short cooling times. The material is effectively at the lower $T$ limit of the
 cooling function at $T\sim 10^4 K$. Outside 10 kpc the cooling rates are now larger than 
 initially, as the initial material at these positions has moved inwards and hotter
 material from the outskirts is falling in. For Model C, at 200 Myr, the central object is still relatively small, 
 which is why its cooling time profile differs from those of Models A and B at this time.

\subsection{Ring formation}
\label{ring}

All of our models exhibit ring structures and a phase dominated by massive clumps.
The ring is very clearly seen in Models A and B, but is significantly less distinct and more obscure in Model C.
The ring formation radius varies from model to model as displayed in the left
panels of Fig. \ref{rings}.

\begin{itemize}

\item  In Model A a shallow peak in the radial surface density distribution is
       produced soon after the start of the simulation at the equilibrium position between the halo gravity and 
       the initial angular momentum of the gas contributing to the 
       first stage of disk formation (t$\lesssim 100 \ \rm{Myr}$).
       New disk material during this phase preferentially settles at the outer region
       of the already existing disk. Thus, the edge of the disk is always relatively
       distinct, resulting in a steep rise of the gravitational potential in this region.
       This in turn focuses the accretion of material to this region
       (cf. Fig. \ref{focus}) forming a peak in the surface density (i.e. a
       ring) in the outer disk, which further increases
       the gravitational attraction and results in a self-amplification of the ring formation process. 
       Due to the growing angular momentum of the infalling
       gas, the radius of the ring wanders outwards from $R\sim 1\ \rm{kpc}$ to
       $R\sim 2\ \rm{kpc}$ at the time of fragmentation ($t\sim 100 \ \rm{Myr}$).
       The top row of Fig. \ref{rings} displays the surface density and the
       gravitational potential as a function of radius at a time shortly 
       before fragmentation. The dominant ring
       corresponds to a local minimum in the gas-dominated gravitational potential
       and therefore the ring is able to attract material from both the inner and outer regions.

\item In Model B a disk with a radius of $R_{\rm{disk}}\sim 5\ \rm{kpc}$ forms by 
      simultaneous accretion of material at all radii $R \leq R_{\rm{disk}}$ leading
      to an initially flat surface density distribution with an edge, with the disk
      being less distinct than in Model A. 
      The edge acts again as a focal point and leads to large gravitational 
      forces attracting material to this region and to a dip
      in the gravitational potential resulting in a ring at $R\sim 4.0\ \rm{kpc}$ 
      as depicted in the middle row of Fig. \ref{rings}.
      The sizes of the initial disk and the ring are determined by 
      the angular momentum profile of the initial
      constant density sphere and the corresponding radii of centrifugal equilibrium.

\item In Model C the disk forms inside-out with accretion during the first 
      phase occurring at a growing range of radii. The 
      gravitational potential of the gas component is monotonically increasing at
      all times all the way from the center to the outer regions.
      Infalling material is attracted to the central regions, 
      which are growing from ongoing vertical accretion
      and gas migrating radially inward. The first phase of 
      gravitational fragmentation occurs in the central region at
      $t\sim290 \ \rm{Myr}$, and unlike Models A and B not in a ring structure.
      In the next phase the influx of new material peaks at larger 
      radii again. The recently accreted gas forms a ring-like structure
      around the already fragmented, irregular, massive gas clumps in the disk within the ring, 
      which dominate the gravitational potential and thus influence the 
      morphology of this less distinct structure (cf. bottom row of Fig. \ref{rings}).

\end{itemize}

\begin{figure}
\centering 
\includegraphics[width=8.5cm]{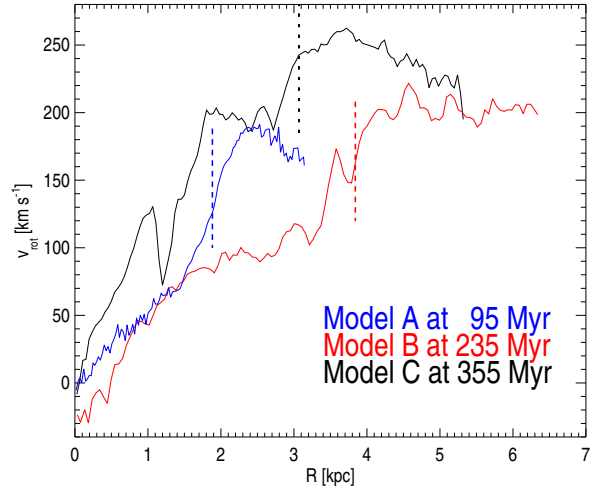}

\caption{The rotation curves for the ring systems of Models A,B,C 
         at the times corresponding to Fig. \ref{rings}. 
         The vertical dashed lines mark the radial positions of the rings.}
\vspace{0.3cm}
\label{RC}
\end{figure}

Fig. \ref{RC} shows the measured gas rotation curves for the three
simulations at the corresponding time depicted in Fig. \ref{rings} with 
the vertical dashed lines indicating the radial positions of the rings. 
Model A shows a monotonically increasing 
rotation curve with a shallower increase in the low surface density inner region and a steeper part in the 
ring region, as one would predict from the corresponding mass profile.
Outside $R\sim2.5 \ \rm{kpc}$ the rotation curve drops indicating 
that no equilibrium disk has yet formed in this region.
The rotation curve for Model B exhibits less monotonic 
behavior indicating a clumpier surface density structure.
The curve is steeper in the center resulting from a small
counter-rotating gas clump, which does not affect the dynamics 
of the rest of the disk. Moving outwards Model B shows similarly to Model A a transition from a
flat inner region at $R \sim 3 \ \rm{kpc}$ to a steep ring-dominated region.
At the time corresponding to Fig. \ref{rings} the disk in Model C has already fragmented. As a result we see
strong local variations in the rotation velocities
produced by massive clumps dominating the gravitational field in their surroundings
(see also \citealp{2004ApJ...611...20I}). Because of the higher
central surface densities model C displays the largest rotation
velocities and the steepest increase in the rotation curve of all the three models.

\begin{figure*}
\centering 
\includegraphics[width=7.5cm]{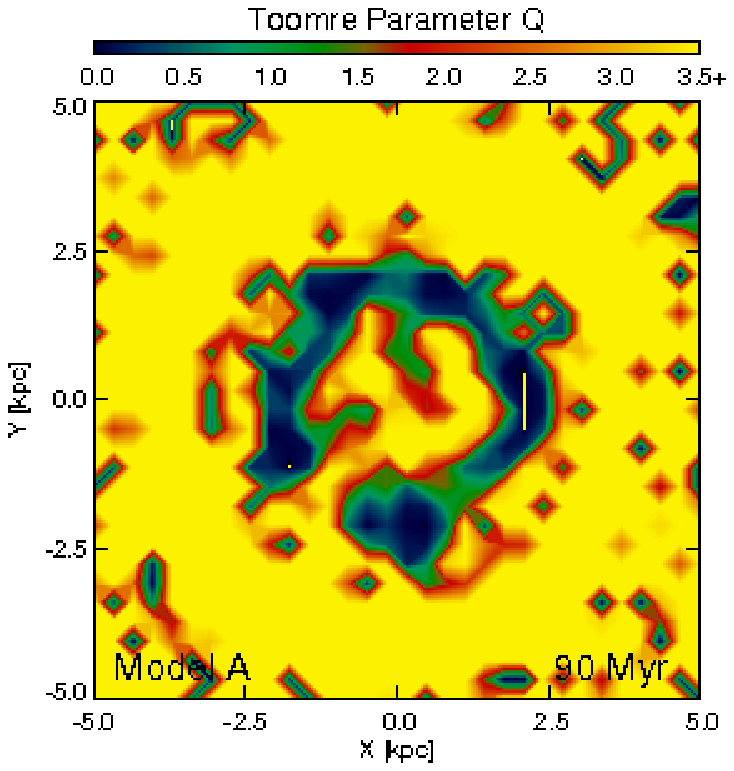}
\includegraphics[width=7.5cm]{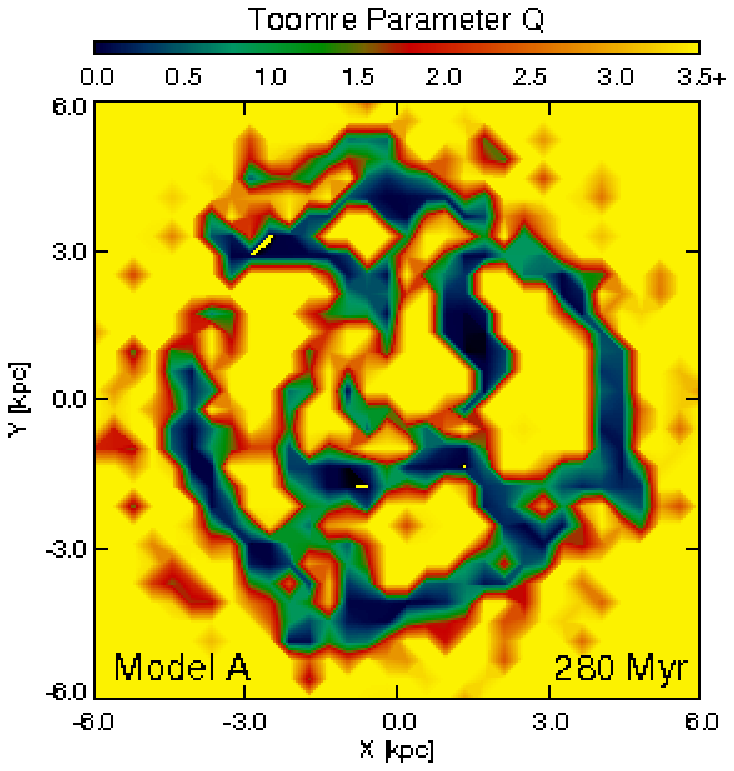}
\caption{Maps of the Toomre  $Q$ parameter for face-on views of the disk region of Model A
         at $t=95 \ \rm Myr$ (ring phase, cf. Fig. \ref{rings}) and 
         at $t=280 \ \rm Myr$ (clump phase, cf. Fig. \ref{observe}).
         Values of $Q>3.5$ are plotted in the same color as $Q=3.5$.
}
\label{Qmap}
\end{figure*}
  
The inclusion of star formation and supernova feedback in Model SF adds an additional source of pressure
and a collisionless component to the simulation. 
The effects of feedback prevent the formation of an initial distinct
ring as seen in Model B, which has identical initial conditions. However, as can be seen in
Fig. \ref{morph} also Model SF develops ring structures, which are short-lived and less distinct than
in Models A and B and more similar to the ring in Model C.

\subsection{Clumps and bulges}
\label{cb}

Fragmentation due to gravitational instability leads to a clump-dominated
evolution phase of our systems. The
Toomre parameter 
\begin{equation}
Q=\frac{\kappa\sigma_R}{\pi G \Sigma},
\end{equation}
predicts instability for $Q<1$, where $\kappa$ is the epicycle frequency,
$\sigma_R$ is the radial velocity dispersion and $\Sigma$ is the surface density,
(\citealp{1964ApJ...139.1217T}, for a revised criterion for turbulent disks see
\citealp{2010arXiv1001.4732R} ).
 
As the temperature of the gas in the disk is typically $T\sim 10^4 \ \rm K$, the corresponding sound speed
is $c_s\sim 10\ \rm kms^{-1}$. As we show in section \ref{sigma}, the velocity dispersion of the gas
is significantly higher than $c_s$ in the region undergoing fragmentation, which justifies the use of $\sigma_R$ for the
calculation of $Q$. The velocity dispersion at this phase is nearly constant with surface density
(see section \ref{sigma}). Thus the surface density distribution $\Sigma$, which evolves
differently as a function of radius in the three models, determines the different fragmentation patterns.
Consequently, fragmentation first occurs in the regions of highest surface
density, which are the ring regions in Models A and B, and the center in Model C.

In the left panel of Fig. \ref{Qmap} we plot a map of the Toomre $Q$ parameter in the disk of Model A in the ring phase at
$t \sim 95 \ \rm Myr$. The ring is beautifully visible as the region with the lowest values of $Q$, with $Q < 1$
indicating that it is unstable. The ring has already started to fragment, as the
shape of the $Q < 1$ region is no longer circular. At larger radii some $Q < 1$ regions are also visible,
however they are outside the equilibrium disk region and the determination of $Q$ is thus not
significant, as the criterion holds only for a thin, differentially rotating disk.

After the fragmentation phase the evolution of the system is dominated by
clumps. We use a friends-of-friends algorithm to detect clumps,
setting a lower mass limit of $M_{\rm min}=10^{7.5}M_{\odot}$, which
corresponds to $N_{\rm min}=158$ SPH particles. The smoothing kernel is
calculated using a number of $N_{\rm{SPH}}=40$ particles and the clumps are thus well resolved.
We also performed tests using the clump detection method described in \citet{2007ApJ...670..237B}, which
defines a clump as a region with a mass above the minimum mass $M_{\rm min}$ and an 
over-density of $\Delta\Sigma\geq 2 \Sigma(R)$, where $\Sigma(R)$ is the average
surface density at the disk radius $R$ and $\Delta\Sigma=\Sigma -\Sigma(R)$. 
Using this clump definition does not alter our conclusions.

The clumps forming in the fragmentation phase have typical initial masses of $M\sim 10^8 M_{\odot}$.
The Jeans mass, $M_j = 2.92  \sigma^3 / (G^{1.5} \rho^{0.5})$ \citep{1987gady.book.....B},
gives an upper mass limit of masses stabilized against gravitational collapse by pressure support.
Using the relations $\Sigma\approx\ 2 rho h$ and $h \approx \sigma_z^2 / (\pi G \Sigma)$, where $h$
is the scale-height of the disk, we find 
\begin{equation}
M_j = 1.0 \times 10^8 M_{\odot} \left(\frac{\sigma_z}{30 \rm {kms^{-1}}}\right)^4  
\left( \frac{\Sigma} {10^{9} M_{\odot}/ \rm{kpc^2}} \right)^{-1}
\end{equation}
Fig. \ref{rings} shows that the typical surface density in rings is $10^{9} M_{\odot}/ \rm kpc^2$.
In section \ref{sigma}, we show that the corresponding vertical velocity dispersion $\sigma_z$ in the ring phase
is 30-40 $\rm kms^{-1}$. Consequently, clumps with masses of $10^8 M_{\odot}$ are consistent with local
Jeans theory. The clumps 
quickly grow in mass due to merging and accretion of diffuse
material. The fraction of gas within the disk region, which is bound in clumps also increases with time
reaching very high values of $f_{\rm clump}\sim 80 \%$ for the gas-only simulations. 

\begin{figure*}
\centering 
\includegraphics[width=15cm]{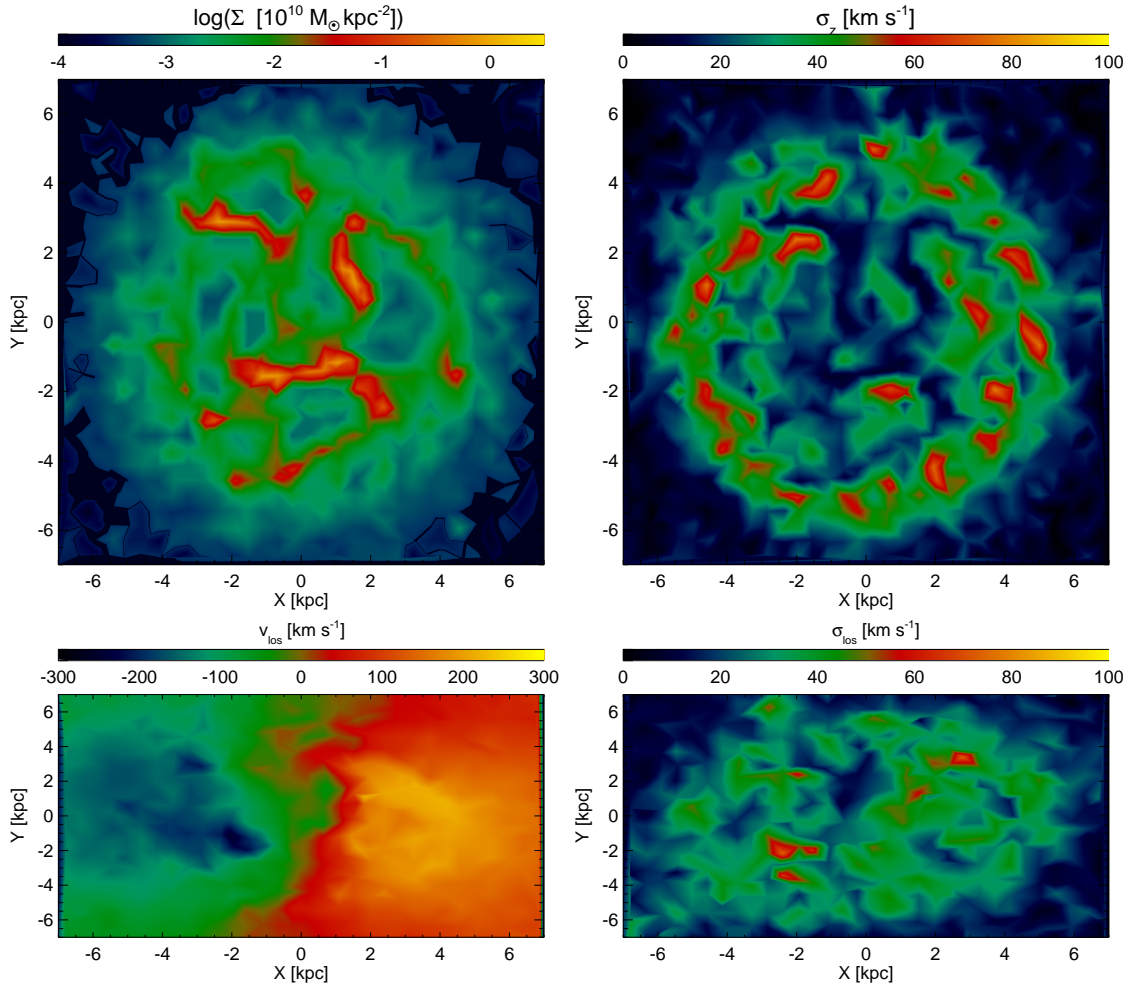}
\vspace{-2cm}
\caption{This figure depicts characteristic quantities of Model A at $t\sim280 \ \rm{Myr}$.
         Top left: Face-on surface density $\Sigma$ map ;
         Top Right: Face-on line-of-sight velocity dispersion $\sigma_z$ map;
         Bottom Left: Line-of sight velocity $v_{\rm{los}}$ map 
                      for an observation with an inclination angle of $i=60^{\circ}$;
         Bottom Right: Line-of sight velocity dispersion $\sigma_{\rm{los}}$ 
                       map for an inclination angle of $i=60^{\circ}$.
}                                                            
\label{observe}
\end{figure*}

The top left panel of Fig. \ref{observe} depicts the surface density map for
Model A at $t\sim280 \ \rm{Myr}$,
which shows four massive clumps ($M\sim10^{9.5}M_{\odot}$) and several smaller
clumps with masses of $M\sim10^{8}M_{\odot}$.
Two of the massive clumps (in the lower half of the plot) are closely interacting.
In the right panel of Fig. \ref{Qmap} we plot a map of the Toomre $Q$ parameter at this time.
The clumps are visible as regions of $Q<1$. The initially rather spherical clumps develop during the simulation into
rotationally-flattened, centrifugally supported mini-disks that interact in the disk plane. The global
Toomre $Q$ parameter is thus no longer an appropriate measure for the stability of the clump regions.
New clumps primarily form at the outer radii of the disk due to newly accreted material.
These outer regions are also Toomre unstable with $Q<1$, as can be seen in the
right panel of Fig. \ref{Qmap}.

The mini-disks are typically aligned in the mid-plane with the exception of Model B, where the fragmentation
occurs during a phase of ongoing strong accretion. The vertical center-of-mass velocity of these clumps
is thus higher, which results in vertical offsets in the position of the clumps. However, also in 
Model B, the vertical velocity 
dispersion within the clumps and their vertical extent are as small as in Models A and C.

\begin{figure*}
\centering 
\includegraphics[width=17cm]{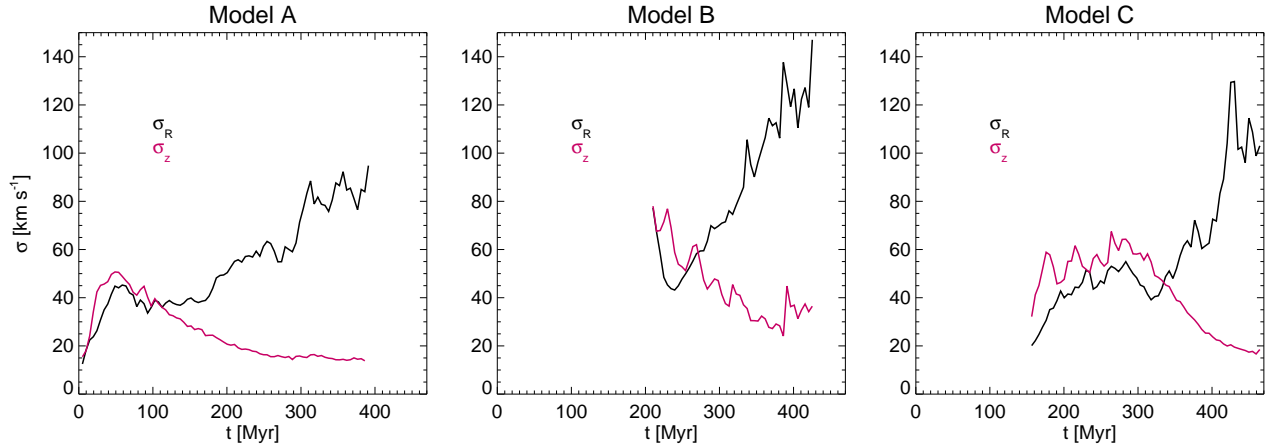}
\caption{The global radial velocity dispersion $\sigma_{R, \rm glob}$ (black lines) and global vertical velocity dispersion
         $\sigma_{z, \rm glob}$ (purple lines) vs time $t$ for the disk-gas in Models A,B and C. For Model A disk formation
         starts instantaneously, whereas for Models B and C it is delayed until $t \sim 200 \ \rm{Myr}$ and
         $t \sim 150 \ \rm{Myr}$ respectively.
}
\label{sigmaVStime}
\end{figure*}

The gas-only models do not show a clear bulge formation process. This is partly caused by the fact
that we did not run these simulations for longer than $t \sim 450 \ \rm{Myr}$, when massive clumps have formed
and we thus expect star formation to play a major role. In all these simulations however the most massive structure
settles in the most central position and can thus be interpreted as a proto-bulge structure. In Models A and C 
the final stages show typically $\sim 5$ clumps with masses $M_{\rm{clump}}\gtrsim 10^9 M_{\odot}$ and $\sim 10$
smaller clumps. Model B shows a high clump-merger rate in the final $\sim 50 \ \rm{Myr}$ resulting in only 3 massive
and three smaller clumps. 
Our gas-only models are thus different from previous models of gravitationally unstable, clump-dominated disks, 
which included prescriptions for star formation (e.g. \citealp{2004A&A...413..547I, 2007ApJ...670..237B}).
These models typically show that massive clumps migrate inwards and coalesce to form a bulge within a few
100 Myr. The main reason for the difference is that almost all of the gas is bound in massive clumps 
and the transfer of angular momentum can only occur between these clumps or between a clump and the 
dark matter component. The dynamical friction timescale is shorter for more massive clumps and thus 
they migrate to central positions more quickly. However, our simulations distinctly lack a diffuse gas component 
which could absorb angular momentum. We have run isolated unstable disk galaxies similar to those of 
\citet{2007ApJ...670..237B} and were able to reproduce the bulge formation process by including 
star formation.

As discussed in section \ref{ring}, the effects of feedback 
significantly alter the formation of over-dense structures within
the disk region of Model SF. The clumps in this model are consequently less dense, less distinct
and less flattened than in Models A,B,C. However, as can be seen in Fig. \ref{morph}
also Model SF develops irregular, clumpy morphology and ring structures. 
The density contrast $\Delta \Sigma_{\rm{tot}} / \Sigma_{\rm{tot}}$, 
where $\Sigma_{\rm{tot}}=\Sigma_{\rm{gas}}+\Sigma_{\rm{stars}}$
of these structures is significantly lower than in the gas-only
simulations. However, if we lower the clump criterion of 
\citet{2007ApJ...670..237B} to $\Delta\Sigma\geq \Sigma(R)$
we still detect clumps of masses $M_{\rm clump}\sim10^{8-9}M_{\odot}$.  
The second row in Fig. \ref{morph} clearly depicts two of these structures 
in the face-on stellar surface density maps.

The morphology of the developing disk changes rapidly on timescales of a few
$t\sim10 \ \rm Myr$. The displayed morphologies show similarities to the models
of \citet{2004A&A...413..547I}. The stellar component and
the larger amount of diffuse gas in this model, allow a more efficient outward
transport of angular momentum and lead to a more efficient inward migration of clumpy material
than in the gas-only simulations, in broad agreement with previous work (e.g. 
\citealp{2004A&A...413..547I, 2007ApJ...670..237B}).
The system of clumps initially displays various irregular morphologies before
forming a bar-like structure ($t\sim 500 \ \rm{Myr}$). The morphology of the bar varies, 
but it persists until the end of the simulation at $t=1.5 \ \rm{Gyr}$ as is depicted 
in the lower rows of Fig. \ref{morph}. At the final stage the model does however not 
depict a central excess of surface density $\Sigma$, as had been found in simulations of isolated, unstable disks 
by \citet{2007ApJ...670..237B}. The radial surface density profile is well-fit by an 
exponential with a scale-length $R_d\approx2.8 \ \rm{kpc}$ out to $R=12 \ \rm{kpc}$.
The final stellar mass of this disk is $M_{\rm{stellar}}\approx 6 \times 10^{10} M_{\odot}$ with a gas fraction
of $\sim 25$ per cent. Initial test runs with static dark potentials and of isolated disks produced stronger 
central components (bulges). This indicates that the interaction of the baryonic 
component with the dark halo plays a strong role in reducing the central dark matter density
(see also \citealp{2009ApJ...697L..38J}). Ongoing accretion
of material with increasing angular momentum onto the disk can also prevent a central surface density excess.

Star formation occurs mainly in the over-dense structures. The morphologies are
very similar for both the gas and stellar components. The SFR averaged over 25 Myr periods,
increases initially from $\sim 30 \; M_{\odot}\rm yr^{-1}$ when the disk starts
forming at $t\sim 220 \ \rm Myr$ to a nearly constant value of 
$\sim 100 \; M_{\odot}\rm yr^{-1}$ between $t\sim 250 \ \rm Myr$ and $t\sim 400 \
\rm Myr$, which is the phase dominated by the irregular disk structures. 
These SFRs are thus similar to observations of high-$z$ disks.
The SFR then
steadily decreases due to gas consumption, being $\sim 60 \; M_{\odot}\rm yr^{-1}$ 
at  $t\sim 500 \ \rm Myr$ , $\sim 30 \; M_{\odot}\rm yr^{-1}$
at $t\sim 1 \ \rm Gyr$  and $\sim 20 \; M_{\odot}\rm yr^{-1}$ at $t\sim 1.5 \ \rm Gyr$.

\subsection{Velocity dispersion}
\label{sigma}

We define a disk region by determining a disk radius $R_{\rm{disk}}$ and a vertical extent $h_z$ selecting
the rotationally supported part of the centrally forming baryonic object.
We then calculate the vertical and radial velocity dispersions $\sigma_{R, \rm glob}$ 
and $\sigma_{z, \rm glob}$ as
\footnote{We use a subscript \textit{glob} to indicate that the dispersion is global,
i.e. calculated for the whole disk region.}
\begin{equation}
\label{defsig}
\sigma_i^2=<v_i^2>-<v_i>^2
\end{equation}
taking into account all gas particles in this region. 
These quantities are thus measures for the global non-circular velocities and disk thickness.
Radial inflow velocities, which vary with position could contribute to $\sigma_{R, \rm glob}$. However,
we do not observe regular inflows in our gas-only Models, which are dominated by clump-clump
interactions after fragmentation.
We plot the dispersions as a function of time in Fig. \ref{sigmaVStime} for the Models A,B and C.
As the temperature of the gas in the disk is typically $T\sim 10^4 \rm K$, the sound speed
is $c_s\sim 10\ \rm kms^{-1}$. As $c_s < \sigma$ is valid in the disk regions, we focus on the
discussion of velocity dispersions.

For Model A (left panel) we find that during the early formation phase
the vertical dispersion is high ($\sigma_{z, \rm glob}\sim 50 \ \rm{kms^{-1}}$) and the radial dispersion stays constant
($\sigma_{R, \rm glob}\sim 40 \ \rm{kms^{-1}}$) until the fragmentation of the ring. The
vertical dispersion, $\sigma_{z, \rm glob}$ starts to decline after $t\sim 50 \ \rm{Myr}$
to a nearly constant level of $\sigma_{z, \rm glob}\sim 15 \ \rm{kms^{-1}}$ after $t\sim 300 \ \rm{Myr}$. 
The radial dispersion, $\sigma_{R, \rm glob}$, on the other hand starts 
to increase with the onset of gravitational instability
reaching values of $\sigma_{R, \rm glob} \sim 80-90 \ \rm{kms^{-1}}$ with strong fluctuations caused by major
clump-clump interactions, which hardly effect $\sigma_{z, \rm glob}$. We ran a test-simulation 
of Model A without gas self-gravity and found that for this
run the z and R dispersions show similar declining behavior, thus confirming that 
disk self-gravity is driving the high radial dispersions.

For Model B (middle panel of Fig. \ref{sigmaVStime}) the initial amplitude of
the velocity dispersion is larger ($\sigma_{\rm glob}\sim80\ \rm{kms^{-1}}$), which is a result
of the almost simultaneous infall of disk material at a large range or radii. After this
first formation phase, $\sigma_{z, \rm glob}$ decreases steeply to reach typical values of 
$\sigma_{z, \rm glob}\sim35 \ \rm{kms^{-1}}$ at $t\sim 400 \ \rm{Myr}$, whereas
$\sigma_{R, \rm glob}$ increases to values of $\sigma_{R, \rm glob} \ \sim100-120\ \rm{kms^{-1}}$,
thus resembling the evolution in Model A.
After showing a clear decreasing tendency, the evolution of $\sigma_{z, \rm glob}$ is affected by a temporary 
increase at $t\sim 390 \ \rm{Myr}$. This increase coincides with two close interactions of two clumps each, which
have an offset in vertical position. This in turn results in vertical gravitational forces increasing the
velocity dispersion temporarily. The merger activity remains high at the end of 
this simulation preventing a further decline in $\sigma_{z, \rm glob}$.

Model C features a longer phase of significant (vertical) accretion onto the inner disk 
regions and an inside-out growth of the disk resulting in a delay of the fragmentation 
phase compared to Model B. The right panel of Fig. \ref{sigmaVStime}
depicting model C thus shows a combination of the features of the other two
Models: $\sigma_{R, \rm glob}$ exhibits first a relatively constant
phase with $\sigma_{R, \rm glob}\sim40 \ \rm{kms^{-1}}$, followed by a steep increase after
the gravitational instability sets in. The vertical dispersion $\sigma_{z, \rm glob}$ is
initially at a high constant level ($\sigma_{z, \rm glob}\sim60 \ \rm{kms^{-1}}$) before 
declining steeply to reach a typical value
of $\sigma_{z, \rm glob}\sim20 \ \rm{kms^{-1}}$ in a similar way as in Model A.

For the corresponding rotation-to dispersion ratios $V/\sigma_{\rm glob}$, we find for radial velocity
dispersion that $2<V/\sigma_{R,\rm glob}<6$ holds at all times in Models A, B and C, which is in agreement with observations
\citep{2009ApJ...697..115C}. For the vertical velocity dispersion, $V/\sigma_{z,\rm glob}$ is only within this interval
during the accretion dominated phase. After that, $V/\sigma_{z,\rm glob}$ increases to values of $\sim 20$, more similar to
low-redshift disks.

The decline in the vertical gas velocity dispersion $\sigma_{z,\rm glob}$ displayed by Models A, B and C results from the
dissipation of energy in random motions $E_{\sigma}$. As has been argued by \citet{2009arXiv0912.0996E}, 
the time evolution of $E_{\sigma}$
is determined by the energy input by accretion (or any other driver of dispersion)
and the dissipation of energy, which acts on a timescale $\tau\sim h/\sigma$, where $h$ is the vertical extent of the disk.
For $h\sim 1 \ \rm kpc$ and $\sigma \sim 50 \ \rm kms^{-1}$, the timescale is $\tau\sim 20 \ \rm Myr$.
The decrease in $\sigma$ in our models is slower, as accretion processes are still active, but become
increasingly less important compared to dissipation. \citet{2009arXiv0912.0996E} argue, that accretion can only
drive high velocity dispersions in an initial phase of disk formation, in agreement with our models.

\begin{figure}
\centering 
\includegraphics[width=8.5cm]{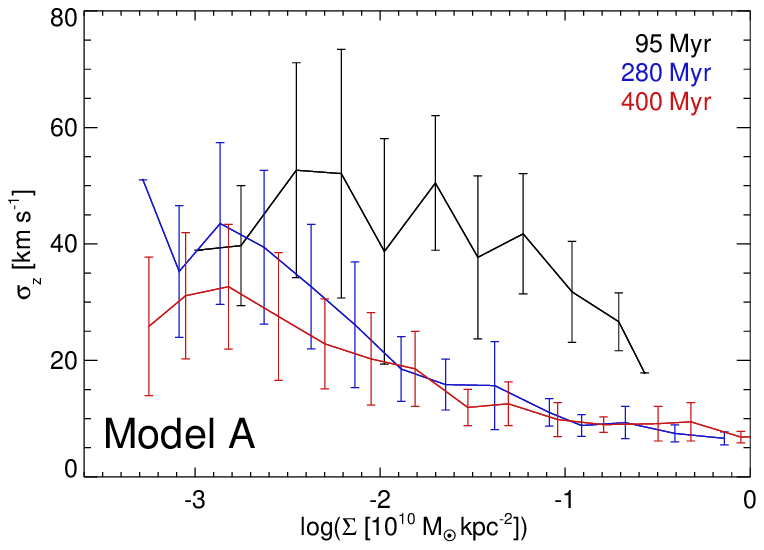}
\includegraphics[width=8.5cm]{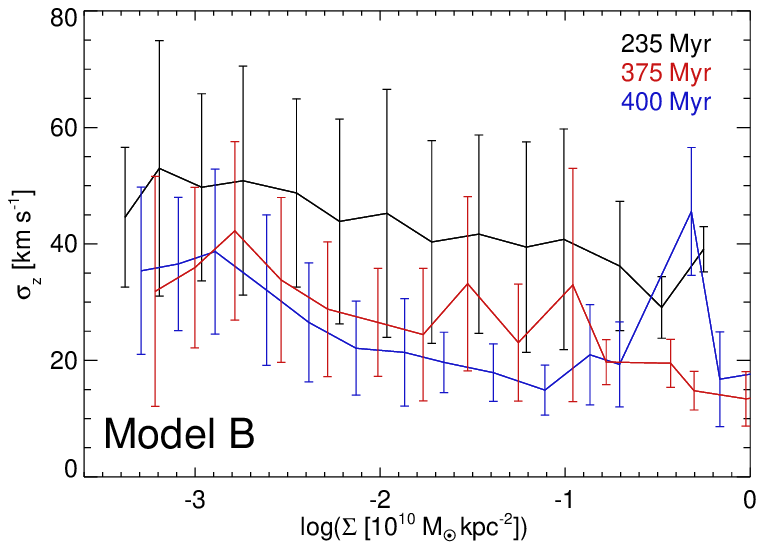}
\includegraphics[width=8.5cm]{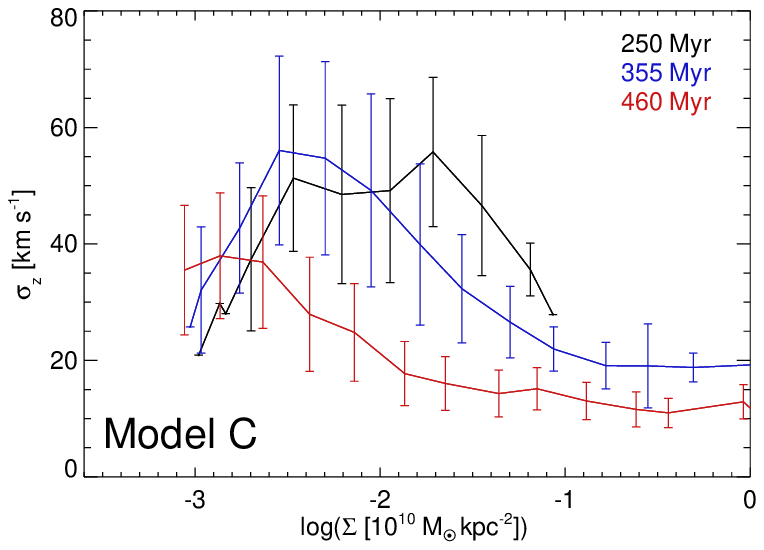}
\caption{Vertical velocity dispersion $\sigma_z$ vs Surface density $\Sigma$. 
         Upper Panel: Model A at 95 Myr (black), 280 Myr (blue) and 400 Myr (red). 
         Middle Panel: Model B at 235 Myr (black), 375 Myr (red) and 400 Myr (blue).
         Lower Panel: Model C at 250 Myr (black), 355 Myr (blue) and 460 Myr (black)}
\label{sigmasigma}
\end{figure}

The analysis above suggests that there should be different velocity dispersion patterns for face-on
and inclined observations of our objects in the massive clump phase.
We show in Fig. \ref{observe} the predicted observational patterns of Model A at 
$t\sim 280 \ \rm{Myr}$, a time at which the expected difference due to the inclination should be large.
We choose Model A as the accretion in the later phases occurs primarily 
at the outer edge of the disk enabling us to distinguish between accretion
and disk-gravity driven velocity dispersion. Furthermore, in gas-only models we do not have 
the additional effects of star formation and feedback complicating the
picture.  The top left panel of Fig. \ref{observe} shows that the system is
dominated by four massive clump complexes, of which the lower two clumps are
interacting. The presence of these clumps also affects the predicted line-of sight velocity 
map for an $i=60^{\circ}$ inclined view as shown in the bottom left panel. 
The top right panel depicts the line-of-sight velocity dispersion map for a face-on and the right 
bottom for an $i=60^{\circ}$ inclined view. Thus, the top right panel shows the vertical 
z-dispersion. High gas surface densities correspond to minima in $\sigma_z$. 
The impact of accretion is reflected by a ring-like structure of high
$\sigma_z$ at large radii. However, on the other hand, for the inclined view the high surface densities
correspond to high values of line-of-sight dispersion, due to the high radial dispersion.
In contrast to the global quantities displayed in figure \ref{sigmaVStime}, these 
velocity dispersions are local quantities, reflecting the substructure in the disks.

We note that not all clumps correspond to high $\sigma_R$ at all times as there are several effects 
playing a role. As discussed in section \ref{cb}, the clumps are centrifugally-supported mini-disks
showing a rotational velocity $v_{\rm{rot}}$ increasing with radius in the center-of-mass frame. 
This variation in $v_{\rm{rot}}$ contributes to the line-of sight dispersion $\sigma_{\rm{los}}$.
However, the clumps are hardly ever axisymmetric (cf. Fig. \ref{observe}), but shaped by clump-clump
interactions and tidal forces, which in each clump create an intrinsic radial velocity dispersion.
Major interactions of clumps such as mergers produce high intrinsic planar dispersions 
thus resulting in the highest values of the inclined-view dispersion.
A third effect contributing to the high values of $\sigma_{R, \rm glob}$ in Fig. \ref{sigmaVStime}
is the clump-to-clump dispersion originating from the different radial center-of-mass velocities
of the individual clumps.

Unlike at $t\sim 280 \ \rm{Myr}$ (Fig. \ref{observe}), where high surface density $\Sigma$ 
is correlated with low vertical velocity dispersion $\sigma_z$, this is not the case for the ring
phase of Model A at $t\approx 95 \ \rm{Myr}$, which is displayed in the top row of Fig. \ref{rings}.
Fig. \ref{sigmaVStime} shows a relatively high  global $\sigma_{z, \rm glob}$ 
at that time. In Fig. \ref{sigmasigma} we
plot disk surface density $\Sigma$ vs local vertical velocity dispersion $\sigma_z$.
The dispersion was determined for a pixelated map of the disk for regions within 
the disk radius $R_{\rm{disk}}$ for Models A,B and C at three different stages of evolution.
The black curve in the upper panel depicts the ring phase of Model A, when dispersion is 
mainly driven by accretion and $\sigma_z$ is relatively independent of $\Sigma$ as 
$\sigma_z\gtrsim 40 \ \rm{km s^{-1}}$ applies to a wide range of surface densities. 
In the regions with the highest surface densities, which correspond to the clumps that are about
to form, the dispersion is already slightly smaller. For 
the subsequent clump-dominated phase, represented by the blue and the red curves, there is a clear 
trend of $\sigma_z$ decreasing with increasing surface density with $\sigma_z\lesssim 20 \ \rm{km s^{-1}}$ 
for the high density regions, a trend that is also depicted in Fig. \ref{observe}.

For Model B, shown in the middle panel, the initial disk/ring phase represented by the black curve
corresponds to high vertical dispersions for all surface densities. The red and the blue curves again 
illustrate general trends of decreasing $\sigma_z$ with time and of lower dispersion for higher 
densities, similar to Model A. However these curves also show features contradicting these trends, 
most prominently the peak of the blue curve at $t\sim 400 \ \rm{Myr}$ for high surface densities.
This distinct feature originates from the clump-clump mergers with vertical offset as discussed
in Section \ref{cb}. The corresponding curves for Model C are depicted in the lower panel. 
They reveal a similar evolution to those of Model A.

The low $z$-dispersions in the massive clumps result from the efficient dissipation of energy
in random motions  $E_{\sigma}$ in the dense gas. It is easier to drive and sustain high 
vertical velocity dispersion by accretion
in low surface density regions than in high $\Sigma$ regions. The reason for this is, that $E_{\sigma}\propto m \sigma^2$
depends strongly on surface density, whereas the energy in the accretion flow per area is independent of surface density.
Moreover, the accretion rates on average decline in all models resulting in a decline in $\sigma_z$ with time
also for low-to medium surface densities.

An anisotropy in velocity dispersions with $\sigma_R>\sigma_z$ for gravitationally unstable disks was already discussed
by \citet{1964ApJ...139.1217T}. \citet{2003MNRAS.344..358B} found $\sigma_R>\sigma_z$ for the gas component
in SPH simulations of (marginally) $Q$-stable, isolated late-type disk galaxies, but attributed this finding to
supernova feedback effects. More recently, \citet{2009MNRAS.392..294A} found a similar anisotropy for HI gas in 
their study of large scale galactic turbulence in simulations of isolated disk galaxies. They note that their
velocity dispersions are 'gravity driven', which includes gravitational instability with $Q<1$, as in our simulations,
but also non-axisymmetric perturbations with $Q>1$.

\begin{figure}
\centering 
\includegraphics[angle=270, width=8.5cm]{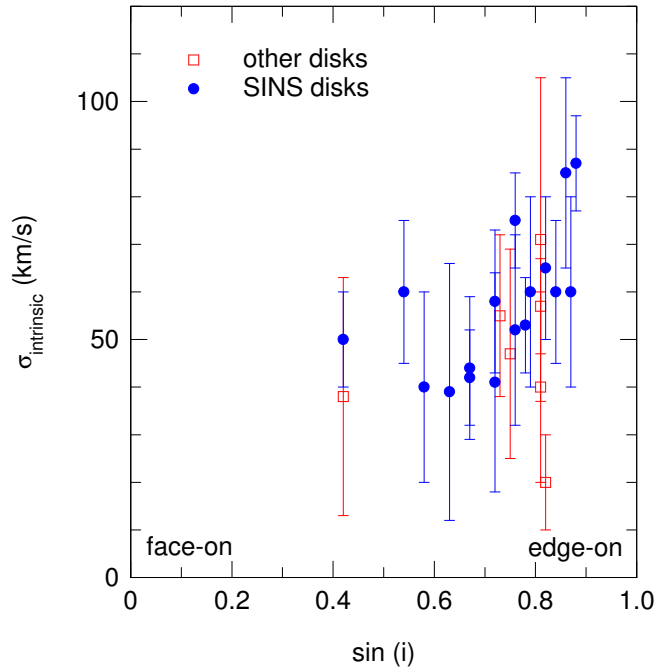}
\caption{Intrinsic velocity dispersion $\sigma_{\rm{intrinsic}}$ vs $\sin(i)$, where $i$ is the 
         inclination angle, under which the disk was observed. Blue dots are for SINS galaxies
         \citep{2009ApJ...706.1364F} and red squares are for other disks (for references see text).
}
\label{IS}
\end{figure}

In Fig. \ref{IS} we plot observed intrinsic velocity dispersions $\sigma_{\rm{intrinsic}}$
in massive $z\sim2$ disk galaxies from the SINS survey \citep{2009ApJ...706.1364F} (blue dots)
and other high-z disks (red squares; \citealp{2009A&A...504..789E,2008Natur.455..775S,2008A&A...488...99V, 2007ApJ...658...78W})
as a function of the inclination angle $i$,
under which the disk has been observed. The observations reveal a systematic
variation of velocity dispersion with $\sin(i)$, where disks that have been observed 
under the lowest inclination angle ($\sin(i)=0.4$), show the lowest dispersions $\sigma_{\rm{intrinsic}}$.
These disks typically have a dispersion of $\sigma_{\rm{intrinsic}}\sim 40 \ \rm{kms^{-1}}$, 
whereas strongly inclined systems show values as high as $\sigma_{\rm{intrinsic}}\sim 90 \ \rm{kms^{-1}}$.
This is similar to our results displayed in Figs. \ref{observe} and \ref{sigmaVStime}, 
where the vertical dispersion $\sigma_{z, \rm glob}$, which is equal to the line-of-sight dispersion
for face-on disks ($\sin(i)=0.0$) is systematically lower than the radial dispersion $\sigma_{R, \rm glob}$, which 
contributes to $\sigma_{\rm{intrinsic}}$ more strongly for more inclined disks.
Unfortunately, no data for face-on disks is yet available, and the line-of-sight velocities of
disks with $\sin(i)=0.4$ still have significant contributions from radial velocities.
The lowest values for $\sigma_{\rm{intrinsic}}$ tend to be slightly higher than the values
for $\sigma_{z, \rm glob}$ in our simulations, with Model B being the best match to these observations.

\begin{figure}
\centering 
\includegraphics[width=8.5cm]{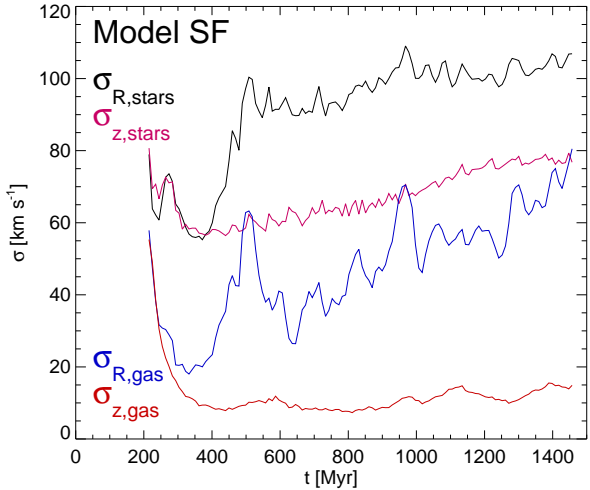}
\caption{Global vertical and radial velocity dispersion $\sigma_{z, \rm glob}$ and $\sigma_{R, \rm glob}$ vs time $t$ s for 
         stars and gas in Model SF.
}
\label{sigmaSF}
\end{figure}

\begin{figure}
\centering 
\includegraphics[width=8.5cm]{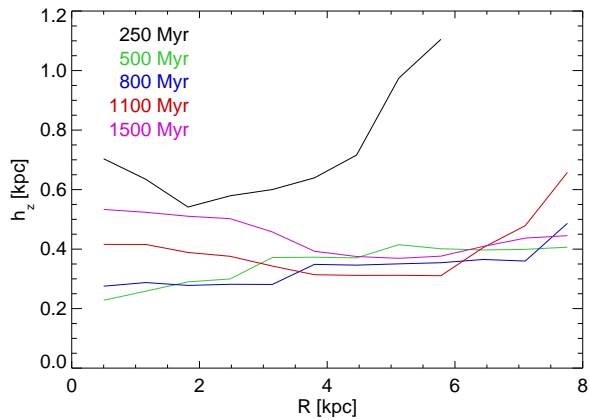}
\caption{Stellar exponential scale heights $h_z$ vs disk radius $R$ for Model SF at times $t=$ 250,
         500,800,1100,1500 Myr.}
\label{heights}
\end{figure}

\subsection{Stellar Disk Scale Heights}

Finally, we study whether small vertical gas velocity dispersions also lead to cold stellar disks
with small scale heights. The global gas velocity dispersion $\sigma_{\rm gas, glob}$ in Model SF,
which is displayed in Fig. \ref{sigmaSF}, initially exhibits
similar behavior to Model B. However, because of the lack of distinct clumps
and clump interactions the increase in the global radial velocity dispersion $\sigma_{R,\rm gas, glob}$ 
is delayed and limited to values of $\sim 40-70 \ \rm{km s^{-1}}$. The vertical
velocity dispersion $\sigma_{z,\rm gas, glob}$ declines similarly to the gas-only models.

We calculate global stellar velocity dispersions $\sigma_{\rm{stars, glob}}$ at a certain time $T$ in Model SF including 
all stellar particles that have formed at $t \leq T$ and show them in Fig. \ref{sigmaSF}. 
Defined as such, the included stellar population
changes from one time $T$ to the next. The stellar velocity dispersion 
$\sigma_{\rm{stars, glob}}$ for Model SF are significantly larger
than the gas velocity dispersions. The vertical stellar dispersion $\sigma_{z,\rm{stars, glob}}$ is 
almost constant, increasing from $\sim 60 \ \rm{km s^{-1}}$ at $t\sim 400 \ \rm{Myr}$
to $\sim 75 \ \rm{km s^{-1}}$ at $t\sim 1.5 \ \rm{Gyr}$. $\sigma_{R,\rm{stars, glob}}$ however
initially ($t\leq 500 \ \rm{Myr}$) shows a similar trend to $\sigma_{R,\rm{gas, glob}}$ 
with the dispersion rising to values of $\sigma_{R,\rm{stars, glob}}\sim 90-100 \ \rm{km s^{-1}}$ 
in the bar phase after a minimum of $\sigma_{R,\rm{stars, glob}}\sim 60 \ \rm{km s^{-1}}$.
The evolution of stellar velocity dispersions $\sigma_{\rm{stars, glob}}$ with time $t$ as displayed in Fig. \ref{sigmaSF}
is thus similar to the results of \citet{2004A&A...413..547I} for central stellar velocity dispersions
in simulations of gravitationally unstable disks.

In Fig. \ref{heights} we display stellar exponential scale-heights $h_z$ as a
function of the disk-radius $R$ extracted from an edge-on view
of the galaxy. In the initial formation phase the stellar disk is thick with
stellar scale heights of $h_z \sim 600-800 \ \rm{pc}$ as can also be seen in the top
row of Fig. \ref{morph}. In this phase the disk is dominated by strong accretion-processes
as discussed in section \ref{sigma}. A population of stars that have formed in clouds
before entering the disk settle in a thick disk. However, new stars typically form near the mid-plane 
of the disk and consequently the scale-heights decrease in the phases with the
highest SFRs. Typical values are $h_z \sim 200-400 \ \rm{pc}$ for $t \sim 500-800 \ \rm{Myr}$, 
comparable to the local Milky Way thin disk. This evolution is in agreement with 
a constant $\sigma_{z,\rm{stars, glob}}$, as the surface density of the disk is continuously 
increasing. After $t \sim 1 \ \rm {Gyr}$ the scale-heights in the central region
begin to increase again up to values $h_z \sim 500 \ \rm{pc}$. This heating is caused
by the bar structure. The scale-height $h_z$ depends only weakly on disk-radius with
the strong early variation at 250 Myr caused by the small radial extent of the disk, which
is in the first stages of its formation. Finally, we note, that the specific prescription for star formation
and the choice of the star formation density threshold $\rho_{\rm th}$, influences
the formation of the disk, especially in the early phases, and thus the disk scale heights.

\section{Discussion}

We have shown that different infall scenarios with disks forming inside-out or simultaneously at
a large range of radii can result in the formation of rings. 
We argue that distinct edges of disks
lead to enhanced accretion of material at these locations which results in rings. Such focal
points could also originate from enhanced infall of material with a certain angular momentum or 
scenarios which have less continuous infall, for example an evolution driven by minor mergers
(\citealp{2009ApJ...699L.178N}, see also \citealp{2004ApJ...616..288B}). 
The position of the ring
is initially determined by the angular momentum distribution of the material, which forms
the first disk in the center of the halo. As displayed by one of our models, the ring radius can increases later-on if material
with higher angular momentum settles into the equatorial plane before the ring breaks up.
We conclude that these ring formation processes should play a role in massive high-redshift ring galaxies.
We also find that rings and clumps characteristically influence the shape of the rotation 
curves of galaxies.

Our simulations indicate that observations of rings, clumps and bulges in disk galaxies at 
redshift $z\sim 2$ are likely to display systems during different stages of evolution. Our 
gas-only models are not able to reproduce the formation of bulges very well, as they lack diffuse
gas and stellar components. However we expect massive star formation to set in once massive
clumps have formed, which would lead to a faster migration of clumps to the center as shown
in Model SF. When we include a simple, but widely-used model for star formation and subsequent 
feedback, our simulations still display the formation of clumpy and ring-like structure, however 
their density-contrasts compared to the rest of the disk are lower than in the gas-only
simulations. Stars mainly form in these structures, which are not long-lived. Our treatment is
too simplistic to allow for a statement on whether feedback can destroy clumps in $z\sim2$ disks 
(cf. \citealp{2010arXiv1001.0765K,2010ApJ...709..191M}).

Interestingly, our Model SF does not show a strong central component in its final stage, its radial
surface density profile is relatively well fit by an exponential with a scale length comparable
to that of the Milky Way. Transfer of angular momentum from the baryonic component to the dark matter
plays an essential role, which is in agreement with the conclusions of \citet{2009arXiv0907.4777B}, who concluded that
adiabatic contraction (e.g. \citealp{1986ApJ...301...27B}) of the halo due to baryonic infall is problematic,
as it predicts halo spin parameters $\lambda$ that are significantly higher than expected. It is also
in agreement with \citet{2010MNRAS.402..776P}, whose simulations of galaxy formation in dark halos
are also not able to reproduce the predictions of the adiabatic contraction theory.

We have run Model SF without feedback and find that due to the lack of additional pressure, 
clumps with similar masses to those of the gas-only runs ($M_{\rm{clump}}\sim 10^{8-9.5} M_{\odot}$), 
but a less flattened morphology, form in the early disk formation phase
($t\sim220-300 \ \rm{Myr}$). The star formation occurs mainly within them at extremely 
high rates ($SFR_{\rm{initial}} \gtrsim 300 \ \rm{M_{\odot}yr^{-1}}$) consuming almost all 
first-infall gas within $\sim 150 \ \rm{Myr}$. After a clump-growth and migration phase 
the clump phase lasts until $t \sim 450 \ \rm{Myr}$. The final system again does not display
a strong central component, the bulge is however more massive than in Model SF.
We note that the observational signatures of typical clumpy $z\sim2$ disks
(SFR, clump fraction) are intermediate to our simulations with and without feedback.

We find that accretion of gas can drive high velocity dispersion of the order of $40-70 \ \rm{km s^{-1}}$
in radial and vertical direction, as long as it penetrates disk regions of low to medium surface densities.
Thus accretion plays an important role in the transient early phases of disk assembly, but affects
the massive clumps region in later evolutionary stages less strongly. 
We note, that the infall of gas clouds produced by thermal instability onto the disks in our simulations
enhances high velocity dispersions. The thermal instability arises due to numerical noise in the
initial conditions, however, as has been argued by \citet{2006MNRAS.370.1612K}, realistic counterparts of
these clouds are likely to exist in gas accretion onto galaxies. The decrease of $\sigma_{z, \rm glob}$ with time
is consistent with estimated dissipation timescales and with predictions of \citet{2009arXiv0912.0996E}
Gravitational instability and the gravitational interactions between massive clumps
however are able to produce high dispersions for in-plane motions. 
The $\sigma_{R, \rm glob}$ vs time plot of Fig. \ref{sigmaVStime} reveals phases of driving velocity dispersion by accretion
and disk-instability, in agreement with the predictions of \citet{2009arXiv0912.0996E}.
Peaks of $\sigma_{R, \rm glob}$ produced by close interactions of the most massive clumps.
Clump-clump mergers increase the radial dispersion for all our models. However, only in Model B, which features
the largest vertical offsets of clumps, they also effect the vertical dispersion due to vertical gravitational
forces between vertically offset interacting clumps. 
Except for phases dominated by mergers of this kind, all models show a characteristic decrease of $\sigma_{z, \rm glob}$
with time $t$, as the dissipation of energy in random motions starts to become increasingly more effective and to dominate
over the driving of high velocity dispersions by accretion.
The rotation-to-radial gas velocity dispersion ratios $V/\sigma_{R, \rm glob}$  
agree very well with observations of high-$z$ disks 
 for our Models A, B and C at all times. For the vertical dispersion, 
$V/\sigma_{z, \rm glob}$ only agrees with observations during the accretion-dominated phase.

In the clump dominated phase, all models also show a characteristic decrease of $\sigma_z$ with 
increasing surface density $\Sigma$ , as long as they are not affected
by mergers of clumps with vertical offsets.
Because of this, face-on views of clumpy disks display minima of the line-of-sight dispersion at the positions
of the clumps, whereas for an inclined view, the clumps correspond to higher than average dispersion
resulting from the high in-plane dispersion. This is not true for the ring phase, in which 
$\sigma_z$ is relatively independent of $\Sigma$. However, already in the phase of initial ring fragmentation,
the clumps, which are about to form, show smaller values of $\sigma_z$ than the rest of the ring.

We present observational evidence for a systematic variation of vertical velocity dispersion with
viewing-angle of the disk for $z\sim2$ galaxies. The observed dispersion for disks with the lowest inclination
angle $sin(i) \sim 0.4$ is still higher than the vertical dispersion in our Models A and C,
only partly explained by the contribution from in-plane motions. 
Model B is a better match to the observations for low-inclination disks.
A possible explanation might be that feedback processes and/or ongoing strong vertical accretion 
into the massive clump region create high radial and vertical velocity dispersions of the order of the observed 
face-on value and vertical offsets of clumps and thus play a significant role vertically, but not
for the in-plane motions, whose dispersion is dominated by disk self-gravity. The symmetry of our
initial conditions also defines a perfect disk plane. Mini-disks
inclined to the disk plane would also give rise to vertical dispersions.

The scale-heights of our forming stellar disk in Model SF are significantly thinner than those 
simulated by \citet{2009ApJ...707L...1B} and do actually not correspond to thick disks. 
Only the early phase, which is characterized by high accretion rates into all disk regions 
displays exponential scale-heights $h_z \sim \ 600-800 \ \rm{pc}$. The final scale-heights 
$h_z\sim 500 \ \rm{pc}$ are larger than that of the Milky Way thin disk but lower than that
of the thick disk. One reason for this is that our Model SF includes feedback and does not 
display distinct, massive and long-lived clumps.  In our test-run without feedback, we find 
that scale-heights are higher by a factor of $\sim 1.5$ due to the more compact clumps
that scatter stars more efficiently, and show stronger radial variations, but still they are
thinner than the Milky Way thick disk. We have also run test-simulations of isolated, massive 
disks with similar initial conditions to \citet{2007ApJ...670..237B} and have been able to 
produce scale-heights of the order of $1 \ \rm{kpc}$. The final scale-height 
depends critically on the Toomre Q factor of the initial disk. The lower Q, the more 
massive and distinct the clumps that form and the more vertically extended the disks.

The scale-heights are however always smaller by a factor of $\sim 2$ compared to \citet{2009ApJ...707L...1B}.
The reason for this might be that Bournaud et al. used a sticky particle code for the gas component,
whereas we use an SPH code. We also find small variations of scale-height with radius, possibly 
on account of different halo potentials. The increase in scale-height is consistent with the idea 
that the disk thickens by stars being scattered off massive clumps. In contrast
to collisionless stellar particles this is not possible for massive gas clouds, which due to their
larger cross-section dissipate kinetic energy in close encounters. A simple estimate for the
velocity a particle can gain perpendicular to its original direction is given by
\begin{equation}
\Delta V= \frac{2GMb}{V_0 (b^2+a^2)},
\end{equation}
where $b$ is the impact parameter, $a$ is the scale-radius of the clump, $V_0$ is the relative velocity of
the clump and the particle and $M$ is the clump mass (\citealp{1987gady.book.....B}, p.474). Taking 
this as an estimate for velocity dispersion and using $M=10^9 M_{\odot}$, $a=1\ \rm{kpc}$, $b=3 \ \rm{kpc}$
and $V_0=50 \ \rm{kms^{-1}}$, we find $\Delta V \sim 50 \ \rm{kms^{-1}}$.
Thus disk heating is likely to occur in clumpy, massive galaxies. However, the efficiency of this process
depends on how compact the clumps are, how long they survive and where and when the stars form. Because
of this it is not clear, whether this is the only process at work in creating thick disks in high-redshift
galaxies and whether they are the birthplace of present-day thick disks.
Distinct, massive clumps are also present in our run of Model SF without feedback. However, the continuous accretion of gas
continuously allows continuous star formation and reduces the disk heating.

Given the limitations inherent in our models, we are able to study essential physical processes
that are important for the early stages of galaxy evolution and their signatures. We are
also able to qualitatively reproduce observational signatures of $z\sim2$ galaxies 
allowing us to draw conclusions about the role these processes play in these objects.
However, more detailed effort is necessary to understand in detail the observational
properties of high-redshift disk galaxies.

%\begin{acknowledgements}

%Any?

%\end{acknowledgements}

\nocite{*}
\bibliography{references}

\begin{thebibliography}{64}
\expandafter\ifx\csname natexlab\endcsname\relax\def\natexlab#1{#1}\fi

\bibitem[{{Agertz} {et~al.}(2009{\natexlab{a}}){Agertz}, {Lake}, {Teyssier},
  {Moore}, {Mayer}, \& {Romeo}}]{2009MNRAS.392..294A}
{Agertz}, O., {Lake}, G., {Teyssier}, R., {Moore}, B., {Mayer}, L., \& {Romeo},
  A.~B. 2009{\natexlab{a}}, \mnras, 392, 294

\bibitem[{{Agertz} {et~al.}(2009{\natexlab{b}}){Agertz}, {Teyssier}, \&
  {Moore}}]{2009MNRAS.397L..64A}
{Agertz}, O., {Teyssier}, R., \& {Moore}, B. 2009{\natexlab{b}}, \mnras, 397,
  L64

\bibitem[{{Aumer} \& {Binney}(2009)}]{2009MNRAS.397.1286A}
{Aumer}, M. \& {Binney}, J.~J. 2009, \mnras, 397, 1286

\bibitem[{{Binney} \& {Tremaine}(1987)}]{1987gady.book.....B}
{Binney}, J. \& {Tremaine}, S. 1987, {Galactic Dynamics, Princeton University
  Press, Princeton, NJ}

\bibitem[{{Blumenthal} {et~al.}(1986){Blumenthal}, {Faber}, {Flores}, \&
  {Primack}}]{1986ApJ...301...27B}
{Blumenthal}, G.~R., {Faber}, S.~M., {Flores}, R., \& {Primack}, J.~R. 1986,
  \apj, 301, 27

\bibitem[{{Bottema}(2003)}]{2003MNRAS.344..358B}
{Bottema}, R. 2003, \mnras, 344, 358

\bibitem[{{Bournaud} \& {Elmegreen}(2009)}]{2009ApJ...694L.158B}
{Bournaud}, F. \& {Elmegreen}, B.~G. 2009, \apjl, 694, L158

\bibitem[{{Bournaud} {et~al.}(2007){Bournaud}, {Elmegreen}, \&
  {Elmegreen}}]{2007ApJ...670..237B}
{Bournaud}, F., {Elmegreen}, B.~G., \& {Elmegreen}, D.~M. 2007, \apj, 670, 237

\bibitem[{{Bournaud} {et~al.}(2009){Bournaud}, {Elmegreen}, \&
  {Martig}}]{2009ApJ...707L...1B}
{Bournaud}, F., {Elmegreen}, B.~G., \& {Martig}, M. 2009, \apjl, 707, L1

\bibitem[{{Bullock} {et~al.}(2001){Bullock}, {Dekel}, {Kolatt}, {Kravtsov},
  {Klypin}, {Porciani}, \& {Primack}}]{2001ApJ...555..240B}
{Bullock}, J.~S., {Dekel}, A., {Kolatt}, T.~S., {Kravtsov}, A.~V., {Klypin},
  A.~A., {Porciani}, C., \& {Primack}, J.~R. 2001, \apj, 555, 240

\bibitem[{{Burkert} {et~al.}(2010){Burkert}, {Genzel}, {Bouche}, {Cresci},
  {Khochfar}, {Sommer-Larsen}, {Sternberg}, {Naab}, {Foerster-Schreiber},
  {Tacconi}, {Shapiro}, {Hicks}, {Lutz}, {Davies}, {Buschkamp}, \&
  {Genel}}]{2009arXiv0907.4777B}
{Burkert}, A., {Genzel}, R., {Bouche}, N., {Cresci}, G., {Khochfar}, S.,
  {Sommer-Larsen}, J., {Sternberg}, A., {Naab}, T., {Foerster-Schreiber}, N.,
  {Tacconi}, L., {Shapiro}, K., {Hicks}, E., {Lutz}, D., {Davies}, R.,
  {Buschkamp}, P., \& {Genel}, S. 2010, ArXiv e-prints, arXiv:0907.4777

\bibitem[{{Burkert} \& {Hartmann}(2004)}]{2004ApJ...616..288B}
{Burkert}, A. \& {Hartmann}, L. 2004, \apj, 616, 288

\bibitem[{{Burkert} \& {Lin}(2000)}]{2000ApJ...537..270B}
{Burkert}, A. \& {Lin}, D.~N.~C. 2000, \apj, 537, 270

\bibitem[{{Ceverino} {et~al.}(2010){Ceverino}, {Dekel}, \&
  {Bournaud}}]{2009arXiv0907.3271C}
{Ceverino}, D., {Dekel}, A., \& {Bournaud}, F. 2010, \mnras, 404, 2151

\bibitem[{{Cresci} {et~al.}(2009){Cresci}, {Hicks}, {Genzel}, {Schreiber},
  {Davies}, {Bouch{\'e}}, {Buschkamp}, {Genel}, {Shapiro}, {Tacconi},
  {Sommer-Larsen}, {Burkert}, {Eisenhauer}, {Gerhard}, {Lutz}, {Naab},
  {Sternberg}, {Cimatti}, {Daddi}, {Erb}, {Kurk}, {Lilly}, {Renzini},
  {Shapley}, {Steidel}, \& {Caputi}}]{2009ApJ...697..115C}
{Cresci}, G., {Hicks}, E.~K.~S., {Genzel}, R., {Schreiber}, N.~M.~F., {Davies},
  R., {Bouch{\'e}}, N., {Buschkamp}, P., {Genel}, S., {Shapiro}, K., {Tacconi},
  L., {Sommer-Larsen}, J., {Burkert}, A., {Eisenhauer}, F., {Gerhard}, O.,
  {Lutz}, D., {Naab}, T., {Sternberg}, A., {Cimatti}, A., {Daddi}, E., {Erb},
  D.~K., {Kurk}, J., {Lilly}, S.~L., {Renzini}, A., {Shapley}, A., {Steidel},
  C.~C., \& {Caputi}, K. 2009, \apj, 697, 115

\bibitem[{{Daddi} {et~al.}(2010){Daddi}, {Bournaud}, {Walter}, {Dannerbauer},
  {Carilli}, {Dickinson}, {Elbaz}, {Morrison}, {Riechers}, {Onodera}, {Salmi},
  {Krips}, \& {Stern}}]{2009arXiv0911.2776D}
{Daddi}, E., {Bournaud}, F., {Walter}, F., {Dannerbauer}, H., {Carilli}, C.~L.,
  {Dickinson}, M., {Elbaz}, D., {Morrison}, G.~E., {Riechers}, D., {Onodera},
  M., {Salmi}, F., {Krips}, M., \& {Stern}, D. 2010, \apj, 713, 686

\bibitem[{{Dekel} {et~al.}(2009{\natexlab{a}}){Dekel}, {Birnboim}, {Engel},
  {Freundlich}, {Goerdt}, {Mumcuoglu}, {Neistein}, {Pichon}, {Teyssier}, \&
  {Zinger}}]{2009Natur.457..451D}
{Dekel}, A., {Birnboim}, Y., {Engel}, G., {Freundlich}, J., {Goerdt}, T.,
  {Mumcuoglu}, M., {Neistein}, E., {Pichon}, C., {Teyssier}, R., \& {Zinger},
  E. 2009{\natexlab{a}}, \nat, 457, 451

\bibitem[{{Dekel} {et~al.}(2009{\natexlab{b}}){Dekel}, {Sari}, \&
  {Ceverino}}]{2009ApJ...703..785D}
{Dekel}, A., {Sari}, R., \& {Ceverino}, D. 2009{\natexlab{b}}, \apj, 703, 785

\bibitem[{{Dib} {et~al.}(2006){Dib}, {Bell}, \&
  {Burkert}}]{2006ApJ...638..797D}
{Dib}, S., {Bell}, E., \& {Burkert}, A. 2006, \apj, 638, 797

\bibitem[{{Elmegreen} \& {Burkert}(2010)}]{2009arXiv0912.0996E}
{Elmegreen}, B.~G. \& {Burkert}, A. 2010, \apj, 712, 294

\bibitem[{{Elmegreen} {et~al.}(2009){Elmegreen}, {Elmegreen}, {Fernandez}, \&
  {Lemonias}}]{2009ApJ...692...12E}
{Elmegreen}, B.~G., {Elmegreen}, D.~M., {Fernandez}, M.~X., \& {Lemonias},
  J.~J. 2009, \apj, 692, 12

\bibitem[{{Epinat} {et~al.}(2009){Epinat}, {Contini}, {Le F{\`e}vre},
  {Vergani}, {Garilli}, {Amram}, {Queyrel}, {Tasca}, \&
  {Tresse}}]{2009A&A...504..789E}
{Epinat}, B., {Contini}, T., {Le F{\`e}vre}, O., {Vergani}, D., {Garilli}, B.,
  {Amram}, P., {Queyrel}, J., {Tasca}, L., \& {Tresse}, L. 2009, \aap, 504, 789

\bibitem[{{F{\"o}rster Schreiber} {et~al.}(2009){F{\"o}rster Schreiber},
  {Genzel}, {Bouch{\'e}}, {Cresci}, {Davies}, {Buschkamp}, {Shapiro},
  {Tacconi}, {Hicks}, {Genel}, {Shapley}, {Erb}, {Steidel}, {Lutz},
  {Eisenhauer}, {Gillessen}, {Sternberg}, {Renzini}, {Cimatti}, {Daddi},
  {Kurk}, {Lilly}, {Kong}, {Lehnert}, {Nesvadba}, {Verma}, {McCracken},
  {Arimoto}, {Mignoli}, \& {Onodera}}]{2009ApJ...706.1364F}
{F{\"o}rster Schreiber}, N.~M., {Genzel}, R., {Bouch{\'e}}, N., {Cresci}, G.,
  {Davies}, R., {Buschkamp}, P., {Shapiro}, K., {Tacconi}, L.~J., {Hicks},
  E.~K.~S., {Genel}, S., {Shapley}, A.~E., {Erb}, D.~K., {Steidel}, C.~C.,
  {Lutz}, D., {Eisenhauer}, F., {Gillessen}, S., {Sternberg}, A., {Renzini},
  A., {Cimatti}, A., {Daddi}, E., {Kurk}, J., {Lilly}, S., {Kong}, X.,
  {Lehnert}, M.~D., {Nesvadba}, N., {Verma}, A., {McCracken}, H., {Arimoto},
  N., {Mignoli}, M., \& {Onodera}, M. 2009, \apj, 706, 1364

\bibitem[{{F{\"o}rster Schreiber} {et~al.}(2006){F{\"o}rster Schreiber},
  {Genzel}, {Lehnert}, {Bouch{\'e}}, {Verma}, {Erb}, {Shapley}, {Steidel},
  {Davies}, {Lutz}, {Nesvadba}, {Tacconi}, {Eisenhauer}, {Abuter}, {Gilbert},
  {Gillessen}, \& {Sternberg}}]{2006ApJ...645.1062F}
{F{\"o}rster Schreiber}, N.~M., {Genzel}, R., {Lehnert}, M.~D., {Bouch{\'e}},
  N., {Verma}, A., {Erb}, D.~K., {Shapley}, A.~E., {Steidel}, C.~C., {Davies},
  R., {Lutz}, D., {Nesvadba}, N., {Tacconi}, L.~J., {Eisenhauer}, F., {Abuter},
  R., {Gilbert}, A., {Gillessen}, S., \& {Sternberg}, A. 2006, \apj, 645, 1062

\bibitem[{{Genzel} {et~al.}(2008){Genzel}, {Burkert}, {Bouch{\'e}}, {Cresci},
  {F{\"o}rster Schreiber}, {Shapley}, {Shapiro}, {Tacconi}, {Buschkamp},
  {Cimatti}, {Daddi}, {Davies}, {Eisenhauer}, {Erb}, {Genel}, {Gerhard},
  {Hicks}, {Lutz}, {Naab}, {Ott}, {Rabien}, {Renzini}, {Steidel}, {Sternberg},
  \& {Lilly}}]{2008ApJ...687...59G}
{Genzel}, R., {Burkert}, A., {Bouch{\'e}}, N., {Cresci}, G., {F{\"o}rster
  Schreiber}, N.~M., {Shapley}, A., {Shapiro}, K., {Tacconi}, L.~J.,
  {Buschkamp}, P., {Cimatti}, A., {Daddi}, E., {Davies}, R., {Eisenhauer}, F.,
  {Erb}, D.~K., {Genel}, S., {Gerhard}, O., {Hicks}, E., {Lutz}, D., {Naab},
  T., {Ott}, T., {Rabien}, S., {Renzini}, A., {Steidel}, C.~C., {Sternberg},
  A., \& {Lilly}, S.~J. 2008, \apj, 687, 59

\bibitem[{{Genzel} {et~al.}(2006){Genzel}, {Tacconi}, {Eisenhauer},
  {F{\"o}rster Schreiber}, {Cimatti}, {Daddi}, {Bouch{\'e}}, {Davies},
  {Lehnert}, {Lutz}, {Nesvadba}, {Verma}, {Abuter}, {Shapiro}, {Sternberg},
  {Renzini}, {Kong}, {Arimoto}, \& {Mignoli}}]{2006Natur.442..786G}
{Genzel}, R., {Tacconi}, L.~J., {Eisenhauer}, F., {F{\"o}rster Schreiber},
  N.~M., {Cimatti}, A., {Daddi}, E., {Bouch{\'e}}, N., {Davies}, R., {Lehnert},
  M.~D., {Lutz}, D., {Nesvadba}, N., {Verma}, A., {Abuter}, R., {Shapiro}, K.,
  {Sternberg}, A., {Renzini}, A., {Kong}, X., {Arimoto}, N., \& {Mignoli}, M.
  2006, \nat, 442, 786

\bibitem[{{Governato} {et~al.}(2010){Governato}, {Brook}, {Mayer}, {Brooks},
  {Rhee}, {Wadsley}, {Jonsson}, {Willman}, {Stinson}, {Quinn}, \&
  {Madau}}]{2010Natur.463..203G}
{Governato}, F., {Brook}, C., {Mayer}, L., {Brooks}, A., {Rhee}, G., {Wadsley},
  J., {Jonsson}, P., {Willman}, B., {Stinson}, G., {Quinn}, T., \& {Madau}, P.
  2010, \nat, 463, 203

\bibitem[{{Haardt} \& {Madau}(1996)}]{1996ApJ...461...20H}
{Haardt}, F. \& {Madau}, P. 1996, \apj, 461, 20

\bibitem[{{Hinshaw} {et~al.}(2009){Hinshaw}, {Weiland}, {Hill}, {Odegard},
  {Larson}, {Bennett}, {Dunkley}, {Gold}, {Greason}, {Jarosik}, {Komatsu},
  {Nolta}, {Page}, {Spergel}, {Wollack}, {Halpern}, {Kogut}, {Limon}, {Meyer},
  {Tucker}, \& {Wright}}]{2009ApJS..180..225H}
{Hinshaw}, G., {Weiland}, J.~L., {Hill}, R.~S., {Odegard}, N., {Larson}, D.,
  {Bennett}, C.~L., {Dunkley}, J., {Gold}, B., {Greason}, M.~R., {Jarosik}, N.,
  {Komatsu}, E., {Nolta}, M.~R., {Page}, L., {Spergel}, D.~N., {Wollack}, E.,
  {Halpern}, M., {Kogut}, A., {Limon}, M., {Meyer}, S.~S., {Tucker}, G.~S., \&
  {Wright}, E.~L. 2009, \apjs, 180, 225

\bibitem[{{Immeli} {et~al.}(2004{\natexlab{a}}){Immeli}, {Samland}, {Gerhard},
  \& {Westera}}]{2004A&A...413..547I}
{Immeli}, A., {Samland}, M., {Gerhard}, O., \& {Westera}, P.
  2004{\natexlab{a}}, \aap, 413, 547

\bibitem[{{Immeli} {et~al.}(2004{\natexlab{b}}){Immeli}, {Samland}, {Westera},
  \& {Gerhard}}]{2004ApJ...611...20I}
{Immeli}, A., {Samland}, M., {Westera}, P., \& {Gerhard}, O.
  2004{\natexlab{b}}, \apj, 611, 20

\bibitem[{{Johansson} \& {Efstathiou}(2006)}]{2006MNRAS.371.1519J}
{Johansson}, P.~H. \& {Efstathiou}, G. 2006, \mnras, 371, 1519

\bibitem[{{Johansson} {et~al.}(2009{\natexlab{a}}){Johansson}, {Naab}, \&
  {Burkert}}]{2009ApJ...690..802J}
{Johansson}, P.~H., {Naab}, T., \& {Burkert}, A. 2009{\natexlab{a}}, \apj, 690,
  802

\bibitem[{{Johansson} {et~al.}(2009{\natexlab{b}}){Johansson}, {Naab}, \&
  {Ostriker}}]{2009ApJ...697L..38J}
{Johansson}, P.~H., {Naab}, T., \& {Ostriker}, J.~P. 2009{\natexlab{b}}, \apjl,
  697, L38

\bibitem[{{Katz} {et~al.}(1996){Katz}, {Weinberg}, \&
  {Hernquist}}]{1996ApJS..105...19K}
{Katz}, N., {Weinberg}, D.~H., \& {Hernquist}, L. 1996, \apjs, 105, 19

\bibitem[{{Kaufmann} {et~al.}(2006){Kaufmann}, {Mayer}, {Wadsley}, {Stadel}, \&
  {Moore}}]{2006MNRAS.370.1612K}
{Kaufmann}, T., {Mayer}, L., {Wadsley}, J., {Stadel}, J., \& {Moore}, B. 2006,
  \mnras, 370, 1612

\bibitem[{{Kaufmann} {et~al.}(2007){Kaufmann}, {Mayer}, {Wadsley}, {Stadel}, \&
  {Moore}}]{2007MNRAS.375...53K}
---. 2007, \mnras, 375, 53

\bibitem[{{Kennicutt}(1998)}]{1998ARA&A..36..189K}
{Kennicutt}, Jr., R.~C. 1998, \araa, 36, 189

\bibitem[{{Kere{\v s}} {et~al.}(2005){Kere{\v s}}, {Katz}, {Weinberg}, \&
  {Dav{\'e}}}]{2005MNRAS.363....2K}
{Kere{\v s}}, D., {Katz}, N., {Weinberg}, D.~H., \& {Dav{\'e}}, R. 2005,
  \mnras, 363, 2

\bibitem[{{Klessen} \& {Hennebelle}(2010)}]{2009arXiv0912.0288K}
{Klessen}, R.~S. \& {Hennebelle}, P. 2010, accepted to A \& A, arXiv:0912.0288

\bibitem[{{Krumholz} \& {Dekel}(2010)}]{2010arXiv1001.0765K}
{Krumholz}, M.~R. \& {Dekel}, A. 2010, \mnras, 635

\bibitem[{{McKee} \& {Ostriker}(1977)}]{1977ApJ...218..148M}
{McKee}, C.~F. \& {Ostriker}, J.~P. 1977, \apj, 218, 148

\bibitem[{{Murray} {et~al.}(2010){Murray}, {Quataert}, \&
  {Thompson}}]{2010ApJ...709..191M}
{Murray}, N., {Quataert}, E., \& {Thompson}, T.~A. 2010, \apj, 709, 191

\bibitem[{{Naab} {et~al.}(2009){Naab}, {Johansson}, \&
  {Ostriker}}]{2009ApJ...699L.178N}
{Naab}, T., {Johansson}, P.~H., \& {Ostriker}, J.~P. 2009, \apjl, 699, L178

\bibitem[{{Navarro} {et~al.}(1996){Navarro}, {Frenk}, \&
  {White}}]{1996ApJ...462..563N}
{Navarro}, J.~F., {Frenk}, C.~S., \& {White}, S.~D.~M. 1996, \apj, 462, 563

\bibitem[{{Noguchi}(1999)}]{1999ApJ...514...77N}
{Noguchi}, M. 1999, \apj, 514, 77

\bibitem[{{Pedrosa} {et~al.}(2010){Pedrosa}, {Tissera}, \&
  {Scannapieco}}]{2010MNRAS.402..776P}
{Pedrosa}, S., {Tissera}, P.~B., \& {Scannapieco}, C. 2010, \mnras, 402, 776

\bibitem[{{Pettini} {et~al.}(1994){Pettini}, {Smith}, {Hunstead}, \&
  {King}}]{1994ApJ...426...79P}
{Pettini}, M., {Smith}, L.~J., {Hunstead}, R.~W., \& {King}, D.~L. 1994, \apj,
  426, 79

\bibitem[{{Rand} \& {Kulkarni}(1990)}]{1990ApJ...349L..43R}
{Rand}, R.~J. \& {Kulkarni}, S.~R. 1990, \apjl, 349, L43

\bibitem[{{Robitaille} \& {Whitney}(2010)}]{2010ApJ...710L..11R}
{Robitaille}, T.~P. \& {Whitney}, B.~A. 2010, \apjl, 710, L11

\bibitem[{{Romeo} {et~al.}(2010){Romeo}, {Burkert}, \&
  {Agertz}}]{2010arXiv1001.4732R}
{Romeo}, A.~B., {Burkert}, A., \& {Agertz}, O. 2010, accepted to MNRAS,
  arXiv:1001.4732

\bibitem[{{Sch{\"o}nrich} \& {Binney}(2009)}]{2009MNRAS.399.1145S}
{Sch{\"o}nrich}, R. \& {Binney}, J. 2009, \mnras, 399, 1145

\bibitem[{{Shapiro} {et~al.}(2008){Shapiro}, {Genzel}, {F{\"o}rster Schreiber},
  {Tacconi}, {Bouch{\'e}}, {Cresci}, {Davies}, {Eisenhauer}, {Johansson},
  {Krajnovi{\'c}}, {Lutz}, {Naab}, {Arimoto}, {Arribas}, {Cimatti}, {Colina},
  {Daddi}, {Daigle}, {Erb}, {Hernandez}, {Kong}, {Mignoli}, {Onodera},
  {Renzini}, {Shapley}, \& {Steidel}}]{2008ApJ...682..231S}
{Shapiro}, K.~L., {Genzel}, R., {F{\"o}rster Schreiber}, N.~M., {Tacconi},
  L.~J., {Bouch{\'e}}, N., {Cresci}, G., {Davies}, R., {Eisenhauer}, F.,
  {Johansson}, P.~H., {Krajnovi{\'c}}, D., {Lutz}, D., {Naab}, T., {Arimoto},
  N., {Arribas}, S., {Cimatti}, A., {Colina}, L., {Daddi}, E., {Daigle}, O.,
  {Erb}, D., {Hernandez}, O., {Kong}, X., {Mignoli}, M., {Onodera}, M.,
  {Renzini}, A., {Shapley}, A., \& {Steidel}, C. 2008, \apj, 682, 231

\bibitem[{{Springel}(2005)}]{2005MNRAS.364.1105S}
{Springel}, V. 2005, \mnras, 364, 1105

\bibitem[{{Springel} {et~al.}(2005){Springel}, {Di Matteo}, \&
  {Hernquist}}]{2005MNRAS.361..776S}
{Springel}, V., {Di Matteo}, T., \& {Hernquist}, L. 2005, \mnras, 361, 776

\bibitem[{{Springel} \& {Hernquist}(2003)}]{2003MNRAS.339..289S}
{Springel}, V. \& {Hernquist}, L. 2003, \mnras, 339, 289

\bibitem[{{Stark} {et~al.}(2008){Stark}, {Swinbank}, {Ellis}, {Dye}, {Smail},
  \& {Richard}}]{2008Natur.455..775S}
{Stark}, D.~P., {Swinbank}, A.~M., {Ellis}, R.~S., {Dye}, S., {Smail}, I.~R.,
  \& {Richard}, J. 2008, \nat, 455, 775

\bibitem[{{Steidel} {et~al.}(2010){Steidel}, {Erb}, {Shapley}, {Pettini},
  {Reddy}, {Bogosavljevi{\'c}}, {Rudie}, \& {Rakic}}]{2010arXiv1003.0679S}
{Steidel}, C.~C., {Erb}, D.~K., {Shapley}, A.~E., {Pettini}, M., {Reddy},
  N.~A., {Bogosavljevi{\'c}}, M., {Rudie}, G.~C., \& {Rakic}, O. 2010, accepted
  to ApJ

\bibitem[{{Sutherland} \& {Dopita}(1993)}]{1993ApJS...88..253S}
{Sutherland}, R.~S. \& {Dopita}, M.~A. 1993, \apjs, 88, 253

\bibitem[{{Tacconi} {et~al.}(2010){Tacconi}, {Genzel}, {Neri}, {Cox}, {Cooper},
  {Shapiro}, {Bolatto}, {Bouch{\'e}}, {Bournaud}, {Burkert}, {Combes},
  {Comerford}, {Davis}, {Schreiber}, {Garcia-Burillo}, {Gracia-Carpio}, {Lutz},
  {Naab}, {Omont}, {Shapley}, {Sternberg}, \& {Weiner}}]{2010arXiv1002.2149T}
{Tacconi}, L.~J., {Genzel}, R., {Neri}, R., {Cox}, P., {Cooper}, M.~C.,
  {Shapiro}, K., {Bolatto}, A., {Bouch{\'e}}, N., {Bournaud}, F., {Burkert},
  A., {Combes}, F., {Comerford}, J., {Davis}, M., {Schreiber}, N.~M.~F.,
  {Garcia-Burillo}, S., {Gracia-Carpio}, J., {Lutz}, D., {Naab}, T., {Omont},
  A., {Shapley}, A., {Sternberg}, A., \& {Weiner}, B. 2010, \nat, 463, 781

\bibitem[{{Toomre}(1964)}]{1964ApJ...139.1217T}
{Toomre}, A. 1964, \apj, 139, 1217

\bibitem[{{van Starkenburg} {et~al.}(2008){van Starkenburg}, {van der Werf},
  {Franx}, {Labb{\'e}}, {Rudnick}, \& {Wuyts}}]{2008A&A...488...99V}
{van Starkenburg}, L., {van der Werf}, P.~P., {Franx}, M., {Labb{\'e}}, I.,
  {Rudnick}, G., \& {Wuyts}, S. 2008, \aap, 488, 99

\bibitem[{{Wright} {et~al.}(2007){Wright}, {Larkin}, {Barczys}, {Erb},
  {Iserlohe}, {Krabbe}, {Law}, {McElwain}, {Quirrenbach}, {Steidel}, \&
  {Weiss}}]{2007ApJ...658...78W}
{Wright}, S.~A., {Larkin}, J.~E., {Barczys}, M., {Erb}, D.~K., {Iserlohe}, C.,
  {Krabbe}, A., {Law}, D.~R., {McElwain}, M.~W., {Quirrenbach}, A., {Steidel},
  C.~C., \& {Weiss}, J. 2007, \apj, 658, 78

\bibitem[{{Zhao} {et~al.}(2009){Zhao}, {Jing}, {Mo}, \&
  {B{\"o}rner}}]{2009ApJ...707..354Z}
{Zhao}, D.~H., {Jing}, Y.~P., {Mo}, H.~J., \& {B{\"o}rner}, G. 2009, \apj, 707,
  354

\end{thebibliography}
\end{document}